\documentclass[12pt]{article}

\usepackage{amsthm,amsmath,amsfonts,amssymb,mathrsfs}
\usepackage[colorlinks,citecolor=blue,urlcolor=blue]{hyperref}
\usepackage{graphicx}
\usepackage{algorithm,algorithmic}
\usepackage{multirow,booktabs,subcaption}
\usepackage{xcolor}
\usepackage{enumerate}
\usepackage[authoryear]{natbib}
\usepackage{url}

\def\bfX{\mathbf{X}}

\def\calR{\mathcal{R}}
\def\GP{\mathcal{GP}}

\def\bfpsi{\boldsymbol{\psi}}
\def\bfb{\mathbf{b}}
\def\bfI{\mathbf{I}}

\def\bftheta{\boldsymbol{\theta}}

\def\mN{\mathrm{N}}
\def\mbR{\mathbb{R}}
\def\bfzero{\boldsymbol{0}}
\def\bfLambda{\boldsymbol{\Lambda}}
\def\diag{\mathrm{diag}}
\def\bfe{\mathbf{e}}

\def\scrN{\mathscr{N}}
\def\bfx{\mathbf{x}}
\def\bfy{\mathbf{y}}
\def\bfW{\mathbf{W}}
\def\bfP{\mathbf{P}}
\def\bfTheta{\boldsymbol{\Theta}}
\def\IG{\mathrm{IG}}
\def\mvec{\mathrm{vec}}
\def\bfD{\mathbf{D}}
\def\cI{\mathcal{I}}
\def\bfq{\mathbf{q}}
\def\bfB{\mathbf{B}}

\def\bfC{\mathbf{C}}

\def\spacingset#1{\renewcommand{\baselinestretch}%
{#1}\small\normalsize} \spacingset{1}

\begin{document}
\title{\bf Bayesian Image-on-Image Regression via Deep Kernel Learning based Gaussian Processes}

\author{Guoxuan Ma$^1$, Bangyao Zhao$^1$, Hasan Abu-Amara$^2$ and Jian Kang$^1$\thanks{To whom correspondence should be addressed: jiankang@umich.edu},\\ Department of Biostatistics, University of Michigan$^1$,\\ Department of Epidemiology, University of Michigan$^2$}
\date{\vspace{-6ex}}

\maketitle

\begin{abstract}
In neuroimaging studies, it becomes increasingly important to study associations between different imaging modalities using image-on-image regression (IIR), which faces challenges in interpretation, statistical inference, and prediction. Our motivating problem is how to predict task-evoked fMRI activity using resting-state fMRI data in the Human Connectome Project (HCP). The main difficulty lies in effectively combining different types of imaging predictors with varying resolutions and spatial domains in IIR. To address these issues, we develop Bayesian Image-on-image Regression via Deep Kernel Learning Gaussian Processes (BIRD-GP) and develop efficient posterior computation methods through Stein variational gradient descent. We demonstrate the advantages of BIRD-GP over state-of-the-art IIR methods using extensive simulations where we synthesize data based on MNIST, Fashion MNIST and fMRI data from HCP. For HCP data analysis using BIRD-GP, we combine the voxel-wise fALFF maps and region-wise connectivity matrices to predict fMRI contrast maps for language and social recognition tasks. We show that fALFF is less predictive than the connectivity matrix for both tasks. Additionally, we identify features from the resting-state fMRI data that are important for task fMRI prediction. 
\end{abstract}

\clearpage

\spacingset{1.5}

\section{Introduction}
In recent large-scale neuroimaging studies, different types of brain images can be collected from the same participants. Typical imaging modalities include structure magnetic resonance imaging (sMRI) and resting-state or task-based functional MRI (fMRI). A question of great interest in multimodal neuroimaging analysis is the study of the associations between different imaging modalities. However, it is unclear how to effectively combine different types of imaging predictors especially those collected with different spatial resolutions or on different domains, e.g., using the region-wise connectivity matrix and voxel-wise imaging statistics from resting-state fMRI data to make predictions on the task-based fMRI contrast maps. 

Task-based fMRI measures brain activity in response to specific tasks of interest. It has been invaluable in investigating the neural mechanisms underlying processes in the human brain. Task-based fMRI finds extensive application in characterizing brain functional anatomy and deriving neural biomarkers for various tasks [\cite{mcnab2008prefrontal, gordon2017precision, wang2019characterizing, ngo2022predicting}]. However, collecting large-scale task fMRI data is costly as it requires careful experimental design and expensive subject training. Conversely, resting-state fMRI data acquisition is comparatively simpler and less prone to confounding factors [\cite{power2014studying, dubois2016building}]. Previous studies have found that the resting-state functional connectivity and the task-evoked brain activity are positively correlated [\cite{HARREWIJN2020116301}]. It also has been shown that the individual variations in task-invoked brain activities can be explained by the brain activities at rest and the brain anatomic structure [\cite{tavor2016task}]. The value of task-based fMRI alongside the accessibility and cost-effectiveness of resting-state fMRI elicits the need of predicting task-based fMRI using resting-state fMRI data. This research question motivates the development of an efficient statistical inference tool: image-on-image regression (IIR), where both the predictor and outcome variables may involve high-resolution images. 

\begin{figure}[t]
\small
\spacingset{1}
\centering
\includegraphics[scale = 0.42]{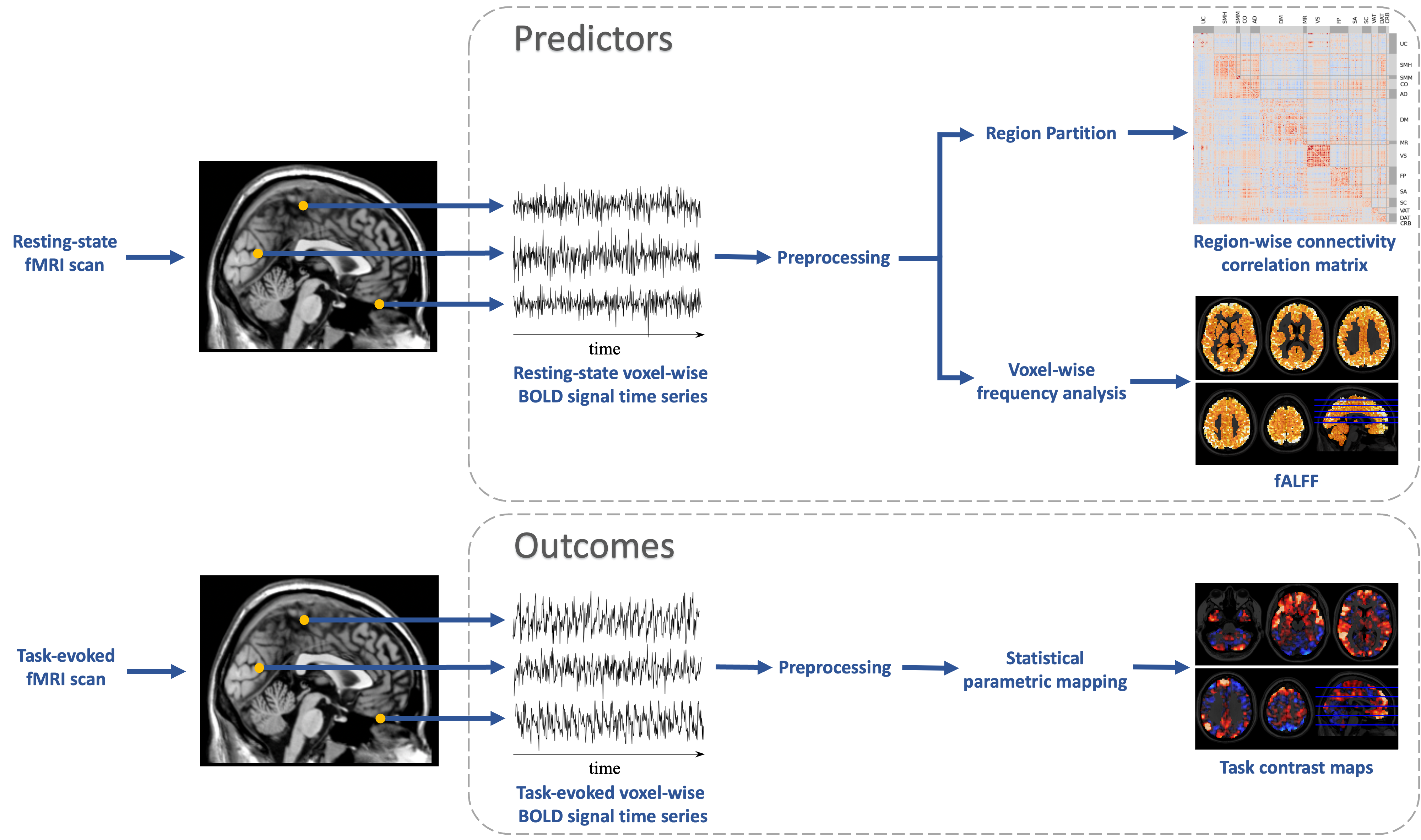}
\caption{HCP fMRI data pipeline description. From raw fMRI scan, both resting-state and task-evoked blood oxygenation level dependent (BOLD) time series are extracted per participant, which then undergo the HCP minimally preprocessed pipeline. The preprocessed resting-state images are regionally summarized, yielding connectivity matrices, while voxel-level frequency analysis generates fALFF maps. The preprocessed task-evoked BOLD signals undergo statistical parametric mapping, generating task-related contrast maps.
}
\label{fig:hcp_data}
\end{figure}

The motivating dataset comes from the Human Connectome 1200 release [\cite{van2012human, HCP2017}], where resting-state and task-evoked fMRI data of multiple modalities are available, including the fractional amplitude of low frequency fluctuation (fALFF), the region-wise connectivity matrix and task fMRI contrast maps. The fALFF is one type of images derived from resting-state fMRI data. It measures the relative contribution of low frequency fluctuations within a specific frequency band to the detectable frequency range of the resting-state fMRI time series [\cite{zou2008improved}], and has been shown to bear predictability to clinical outcomes [\cite{zhao2015bayesian, egorova2017fractional}]. The resting-state connectivity correlation matrix is derived from four time-course files collected from two different fMRI sessions, each comprising 264 nodes. The task-evoked fMRI contrast maps are derived by statistical parametric mapping for the preprocessed voxel-wise time series from fMRI scan during task time. In our study, we perform separate regression analyses of contrast maps from two tasks, i.e., the story-math contrast from the language task and the random-baseline contrast from the social recognition task, while using the resting-state fALFF and region-wise connectivity correlation matrix as predictors. Figure \ref{fig:hcp_data} provides a description of HCP data analyzed in our study. We aim to investigate whether the fALFF map and the region-wise connectivity matrix are capable to predict either task contrast maps, and which brain regions are the most predictable in these task contrast maps by the predictors.

Deep convolutional neural networks are widely used in computer vision applications for image prediction. These methods preserve spatial information within the image through convolutional layers. Most algorithms using deep convolutional neural networks are developed for specific tasks [\cite{santhanam2017generalized}] and commonly utilize architectures including the visual geometry group (VGG) network and the residual neural network (ResNet). However, these architectures often incorporate extraneous information, introducing biases into predicted images [\cite{isola2017image}]. To address general image-to-image regression problems, Recursively Branched Deconvolutional Network (RBDN) has been proposed. This method extracts features to create a composite map which is processed through multiple convolutional layers for each task [\cite{santhanam2017generalized}]. Despite its versatility, RBDN requires input and output images to have the same size, making it unsuitable for some tasks, and it uses the mean-squared error as the objective function which may be limited for some neuroimaging applications [\cite{isola2017image}] such as the prediction of fMRI images.

Generative adversarial networks (GANs) have also been applied to image-to-image regression tasks. In general, GANs address the challenge of lacking a widely-accepted objective function for such tasks by employing a competitive framework, where neural networks are employed to work adversarially to improve image prediction accuracy [\cite{isola2017image}]. However, a limitation of GANs is their inability to directly link the predicted image to input image [\cite{DBLP:journals/corr/abs-1803-04469}]. This issue can be mitigated by using conditional GANs which incorporate additional information to map random noises to the predicted image [\cite{isola2017image}]. One prominent method, Pix2Pix [\cite{isola2017image}], generates sharper images by combining adversarial loss with reconstruction loss. The reconstruction loss is defined as the $L_1$ distance rather than the Euclidean ($L_2$) distance between the generated image and the ground truth [\cite{DBLP:journals/corr/abs-1803-04469}]. Other approaches improve predictions by coupling GAN with a discriminator such as a variational autoencoder (VAE) [\cite{DBLP:journals/corr/abs-1803-04469}]. However, GANs face challenges related to their objective functions. For instance, technical issues such as achieving Nash equilibrium between competing neurons can prevent meaningful predictions [\cite{DBLP:journals/corr/abs-1803-04469}]. Moreover, the saturation of the objective function can lead to a zero gradient, impeding model learning in certain cases [\cite{DBLP:journals/corr/abs-1803-04469}]. GANs also require substantial effort to train properly. Furthermore, the absence of a specific, interpretable objective function may limit the identification of features relevant to predictions, undermining the scientific utility of GANs in tasks like resting-state fMRI studies
 
Interpretability is of particular importance to machine learning techniques [\cite{Carvalho2019}]. For methods to be useful in scientific inquiries on task-rest experiments, the models should provide insight into how task image predictions are derived from rest images. While there is existing literature on interpretable GANs [\cite{NEURIPS2020_6fe43269}] and convolutional neural networks (CNNs) [\cite{Zhang2018}], there is a notable lack of research on interpretable IIR methods in machine learning. This gap limits the applicability of these methods for our motivating neuroimaging applications.

Statistical inference on IIR for neuroimaging studies is more interpretable but is challenging due to the high dimensionality of model parameters and the heterogeneity in activation patterns among individuals. The spatial dependence or correlations among predictors and outcomes can be complex and hard to quantify. In some studies, IIR cannot produce very accurate predictions due to the low signal to noise ratio and relatively small sample sizes. In addition to neuroimaging applications, IIR has attracted growing scientific interests in many other fields such as spatial economics [\cite{gelfand2003spatial}], genomics [\cite{morris2011automated}] and computer vision [\cite{santhanam2017generalized}], where the modeling, inferences and predictions also face similar challenges. 

Various IIR methods have been developed, each motivated by different applications. A simple linear regression model [\cite{tavor2016task}] has been proposed to make predictions on the task-based fMRI data using the resting fMRI and structural MRI as predictors. This linear regression method assumes the predictor and outcome images are collected in the same imaging space and partitions the imaging space into multiple subregions. For each subject in the training data, a linear regression model is fitted over voxels within each region. The average model fits from the training data are used to predict the task activity of a new subject. This method performs well for small datasets while it lacks the flexibility to capture complex associations between the predictor and outcome images and ignores spatial correlations among voxels.

The spatial Bayesian latent factor model (SBLF) for IIR [\cite{Guo2022}] has been recently proposed and has shown successful applications in neuroimaging studies. SBLF introduces the spatial latent factors to establish connections between the outcome images and predictor images. It explicitly accounts for the spatial dependence among voxels in the images, resulting in improved prediction accuracy fMRI data analysis. However, SBLF may suffer the over-fitting issues due to the inclusion of both individual specific and basis function specific random effects. In addition, the posterior computation of SBLF is extremely challenging and the current algorithm is inefficient and not scalable for analyzing the large-scale imaging data.

To address the challenges of IIR and the limitations of the current methods for neuroimaging applications, in this article, we develop a Bayesian Image-on-image Regression via Deep kernel learning based Gaussian Processes (BIRD-GP). BIRD-GP is a new Bayesian hierarchical model for IIR by integrating deep neural networks (DNN) and Gaussian processes (GP) with kernel learning. It is a framework with two-stage analysis: the image projection via the basis expansion approach (Stage 1) and the nonlinear regression via DNN (Stage 2). This framework substantially reduces the number of model parameters compared to other deep learning methods but it still has the flexibility to capture complex associations between the predictor and outcome images, leading to interpretable model fitting and accurate prediction performances. We propose a novel method to learn the covariance kernel or equivalently the orthonormal basis functions of the GPs via DNN. BIRD-GP can capture detailed characteristics of the predictor and outcome images, substantially facilitate the statistical efficiency in estimating the model parameters, and explicitly provide a set of basis images that can greatly improve interpretability. Under the Bayesian framework, BIRD-GP can also produce valid prediction uncertainty measures via the posterior probability. For posterior computation, we develop a hybrid posterior computation algorithm by combining the Gibbs sampler and the Stein variational gradient descent method. It is computationally efficient and scalable to large scale neuroimaging data. It also can be straightforwardly implemented in parallel. 

We conduct extensive experiments to evaluate the performance of BIRD-GP, employing synthetic datasets based on MNIST, Fashion MNIST and fMRI data from HCP. BIRD-GP outperformed all competing methods in these experiments. Furthermore, we apply BIRD-GP to two IIR tasks to regress the language task story-math contrast maps and the social recognition random-baseline contrast maps using fALFF images and connectivity correlation matrix from the HCP 1200 release [\cite{HCP2017}]. Results show that BIRD-GP achieve better prediction accuracy than competing methods. Our findings reveal that the language task story-math contrast maps are more predictable than the social recognition random-baseline contrast maps when using fALFF and connectivity as predictors. Furthermore, we observe that connectivity alone demonstrated superior predictability compared to fALFF alone in both tasks. To gain further insights, we visualize and analyze the basis images generated by BIRD-GP using different modalities as predictors for both tasks. 

The rest of the article is structured as follows. In Section 2, we first provide the model formulation in Section~\ref{sec:method_model} and develop the framework of projected predictor image importance calculation in Section~\ref{sec:method_imprtance}, followed by an equivalent model representation in Section~\ref{sec:method_equivalent}. Then, in Section~\ref{sec:metheod_kernel} we describe a novel approach for kernel learning via Deep Neural Networks (DNNs). We discuss the prior specifications in Section~\ref{sec:method_prior}. In Section~\ref{sec:posterior}, we describe the posterior algorithm. We present the results of BIRD-GP and competing methods on synthetic datasets in Section~\ref{sec:synthetic}. We analyze the HCP fMRI data in Section~\ref{sec:hcp}, and conclude the paper in Section \ref{sec:conclusion}.

\section{Method}
Suppose the data consists of $n$ observations of the predictor and outcome images. Let $d_x$ and $d_y$ be the dimension of voxels (or pixels) for the predictor and outcome images, respectively. Let $\calR_x\subset \mathbb{R}^{d_x}$ and $\calR_y \subset\mathbb{R}^{d_y}$ be the collections of voxels to measure image intensities accordingly. For each observation $i$ $(i = 1,\ldots, n)$, let $X_i(v)$ represent the intensity at voxel $v\in \calR_x$ and $Y_i(u)$ represent the intensity at voxel $u\in\calR_y$, respectively. 

\begin{figure}[!ht]
    \small
    \spacingset{1}
    \centering
    \includegraphics[width=0.9\textwidth]{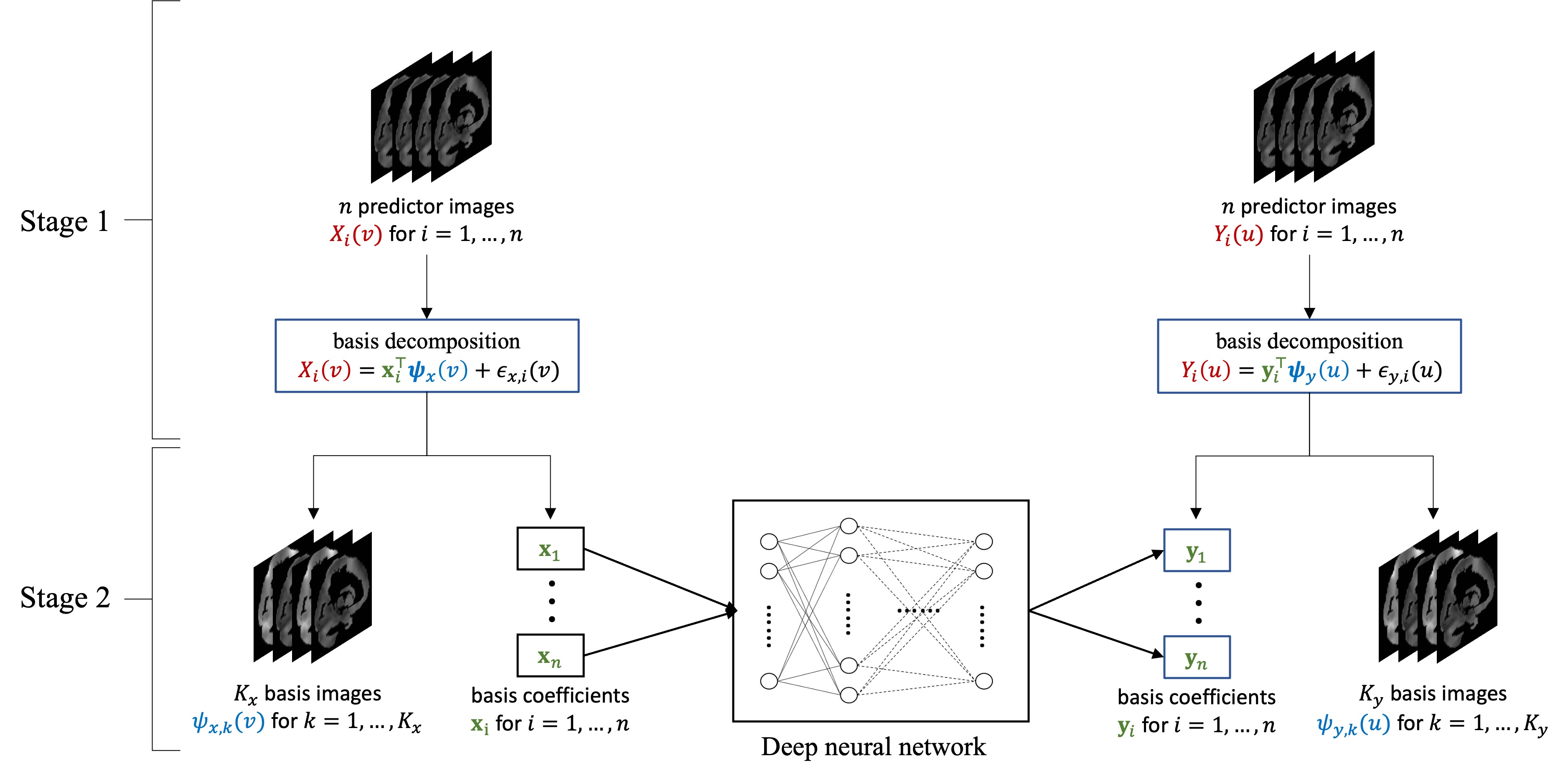}
    \caption{The illustration of BIRD-GP which consists of two stages of analysis. Stage 1: image projection models with the basis expansion approach. Stage 2: deep neural network model for complex associations between projected predictor images and outcome images. In Stage 1, we project images via a basis decomposition approach for predictors and outcome images respectively. The basis functions $\bfpsi_x$ and $\bfpsi_y$ can be learned in a data adaptive manner (detailed in Section \ref{sec:metheod_kernel}) or based on a prespecified kernel function. Then, in Stage 2, we train a deep neural network between the projected predictor images (i.e., basis coefficients for predictors) and the projected outcome images (i.e., basis coefficients for outcomes) for prediction. Under i.i.d. Gaussian errors and Gaussian prior specification on predictor basis coefficients $\bfx_i$, the predictor images $X_i(v)$ follow Gaussian process priors and the outcome images $Y_i(u)$ also follow Gaussian processes conditional on the predictor images (see Section \ref{sec:method_model}).}
    \label{fig:model}
\end{figure}

\subsection{Bayesian image-on-image regression}\label{sec:method_model}

To model the associations between the predictor and outcome images, we consider a two-stage Bayesian method illustrated in Figure \ref{fig:model}. In Stage 1, we model both the predictor and outcome images by a basis expansion approach. We will show our model implies both the predictor and outcome images are realizations of GPs in Section \ref{sec:method_equivalent}.
%Let $K_x$ and $K_y$ be positive integers. 
Let $\bfpsi_x(\cdot) = \{\psi_{x,1}(\cdot),\ldots, \psi_{x,K_x}(\cdot)\}^\top: \calR_x \mapsto \mathbb{R}^{K_x}$ be a vector of $K_x$ orthonormal basis functions for the predictor images, and $\bfpsi_y(\cdot) = \{\psi_{y,1}(\cdot),\ldots,\psi_{y,K_y}(\cdot)\}^\top: \calR_y \mapsto \mathbb{R}^{K_y}$ be a vector of $K_y$ orthonormal basis functions for the outcome images. We have $\sum_{v\in\calR_x} \bfpsi_x(v)\bfpsi_x^\top(v) = \bfI_{K_x}$ and $\sum_{u\in\calR_y} \bfpsi_y(u)\bfpsi_y^\top(u) = \bfI_{K_y}$ where $\bfI_{K_x}$ is an identity matrix with dimensions $K_x \times K_x$. We assume
\begin{align}
X_i(v) &= \bfx_{i}^\top\bfpsi_x(v) + \epsilon_{x,i}(v), \quad \epsilon_{x,i}(v) \sim \mathrm{N}(0,\sigma^2_{x,i}),\label{eq:predictor_img}\\
Y_i(u) &= \bfy_{i}^\top\bfpsi_y(u) + \epsilon_{y,i}(u),\quad \epsilon_{y,i}(u) \sim \mathrm{N}(0,\sigma^2_{y,i}), \label{eq:outcome_img} 
\end{align}
where $\bfx_i \in \mathbb{R}^{K_x}$ and $\bfy_i \in \mathbb{R}^{K_y}$ represent the $i$th projected predictor image and the $i$th projected outcome image in the corresponding Euclidean vector spaces, respectively. The random noises $\epsilon_{x,i}(v)$ and $\epsilon_{y,i}(u)$ explain the variations of observed images that cannot be explained by the basis functions. We assume that $\epsilon_{x,i}(v)$ and $\epsilon_{y,i}(u)$ are mutually independent across $i$, $v$ and $u$. The variances $\sigma_{x,i}^2$ and $\sigma^2_{y,i}$ are image-specific and can be different across different images to accommodate the heterogeneity in noises.
In Stage 2, we specify the joint distributions of the projected predictor images $\bfx_i$ and the projected outcome images $\bfy_i$. We adopt a feed-forward deep neural network (DNN) to model their complex associations,
\begin{align}\label{eq:dnn_link}
\bfx_i &\sim \mN(\bfzero_{K_x},\bfLambda_x),\quad (\bfy_i \mid \bfx_i)\sim \mN\{\scrN(\bfx_i; \bftheta_{xy}),\bfLambda_y\},
 \end{align}
where $\bfzero_{K_x}$ is an all-zero vector of length $K_x$; covariances $\bfLambda_x = \diag\{\lambda_{x,1},..., \lambda_{x,K_x}\}$ and $\bfLambda_y = \diag\{\lambda_{y,1},..., \lambda_{y,K_y}\}$ are diagonal matrices with positive elements approximating the eigenvalues of the GPs for the predictor and outcome images, respectively. The conditional expectation of $\bfy_i$ given $\bfx_i$ is modeled as an $L$-layer feed-forward DNN $\scrN(\cdot\,;\bftheta_{xy}):\mbR^{K_x} \mapsto \mbR^{K_y}$, where each layer has $S_{xy,\ell}$ hidden units ($\ell = 0, 1,\ldots, L$). The output layer dimension $S_{xy,L} = K_y$ and the input layer dimension $S_{xy,0} = K_x$, and $\bftheta_{xy} = \{\bfW_{xy,\ell}, \bfb_{xy,\ell}\}_{\ell=1}^L$ with $\bfW_{xy,\ell}\in \mbR^{S_{xy,\ell}\times S_{xy,\ell-1}}$ and $\bfb_{xy,\ell}\in \mbR^{S_{xy,\ell}}$. 

\subsection{Projected predictor image importance}\label{sec:method_imprtance}
To determine the importance of each element of the projected predictor image $\bfx$, we consider the log density of the outcome $\bfy$ given the predictor $\bfx$, i.e. $\log p(\bfy \mid \bfx, \bftheta_{xy}, \bfLambda_y)$, and define an importance function by the expectation of its derivative
\begin{align}
    \bfq(\bfx, \bfy) = \mathbb{E}_{(\bftheta_{xy}, \bfLambda_y) \sim p(\bftheta_{xy}, \bfLambda_y \mid \mathcal{D})} \frac{\partial}{\partial \bfx} \log p(\bfy \mid \bfx, \bftheta_{xy}, \bfLambda_y),
\end{align}
where the derivative and expectation is exchangeable and $\bfq: \mathbb{R}^{K_x} \times \mathbb{R}^{K_y} \mapsto \mathbb{R}^{K_x}$ is a function of $\bfx$ and $\bfy$ in general and its output dimension is the same as the dimension of $\bfx$. We would like to note that $\bfq(\bfx, \bfy)$ provides a lower bound of the derivative of the log of the posterior predictive distribution, i.e., $\log p(\bfy \mid \bfx, \mathcal{D})$, w.r.t. $\bfx$ by the Jensen's inequality, where $\mathcal{D}$ denotes training data of projected predictor and outcome images. This connection is an analogy of the log marginal density of data and evidence lower bound (ELBO) in the variational inference context. Then we define the importance of the projected predictor image by taking the expectation of the absolute value of $\bfq(\bfx,\bfy)$, i.e., 
\begin{align}\label{eq:basis_importance}
    \textbf{IM} = \mathbb{E}\left| \bfq(\bfx,\bfy) \right| = \int \left| \bfq(\bfx,\bfy)\right| p(\bfx,\bfy \mid \mathcal{D})\; d \bfx d \bfy,
\end{align}
where $|\cdot|$ is an element-wise absolute value function. The expectation is taken with respective to the joint predictive distribution of $\bfx$ and $\bfy$, i.e. $p(\bfx, \bfy \mid \mathcal{D}) = p(\bfy\mid \bfx,\mathcal{D})p(\bfx \mid \mathcal{D})$. Note that \textbf{IM} is a vector of length $K_x$, where each element is the importance measure of the corresponding dimension of the projected predictor image $\bfx$ for prediction. 

The expected magnitude of the partial derivative of the predictive distribution of $\bfy$ with respect to $\bfx$ reflects the strength of association between $\bfx$ and $\bfy$. As a simple example, when $\bfx$ and $\bfy$ are independent, the partial derivative is zero. As another illustration, suppose $(\bfy \mid \bfx)\sim \mN(\bftheta_{xy}\bfx, \bfLambda_y)$ where $\bfx$, $\bfy$, $\bftheta_{xy}$ and $\bfLambda_y$ are all scalars. We assume further that the posterior distribution of $\bftheta_{xy}$ and $\bfLambda_y$ are degenerate, so that $\bftheta_{xy}$ can only take $\beta \in \mathbb{R}$ and $\bfLambda_y$ can only take $\lambda_y \in \mathbb{R}^{+}$. Then, the importance function $\bfq(\bfx, \bfy) = \frac{\partial}{\partial \bfx} \log \mN(\bfy \mid \beta\bfx, \lambda_y)$ and the importance measure of $\bfx$ is $\textbf{IM}=\mathbb{E}_{\bfx, \bfy}\left|\frac{\partial}{\partial \bfx} \log \mN(\bfy \mid \beta\bfx, \lambda_y) \right| = \frac{|\beta|}{\lambda_y}\mathbb{E}_{\bfx, \bfy} \left|\bfy - \beta\bfx\right| = \frac{|\beta|}{\lambda_y}\mathbb{E}_{\bfx} \mathbb{E}_{\bfy\mid\bfx} \left|\bfy - \beta\bfx\right|=\frac{|\beta|}{\lambda_y}\mathbb{E}_{\bfx}\lambda_y\sqrt{\frac{2}{\pi}} = \sqrt{\frac{2}{\pi}}|\beta|$ by noting that $\mathbb{E}_{\bfy\mid\bfx} \left|\bfy - \beta\bfx\right|$ is the expectation of a half-normal random variable. With a linear model, the importance of $\bfx$ is its coefficient magnitude scaled by a factor. 

In practice, the closed-form representation of \eqref{eq:basis_importance} may not be available as in the linear model example, but we can approximate \textbf{IM} via the Monte Carlo method. Suppose we have $S$ samples of $\bftheta_{xy}$ and $\bfLambda_y$ drawn from their posterior distributions, denoted as $\{(\bftheta^{(s)}_{xy},\bfLambda^{(s)}_y)\}_{s=1}^S$, and denote $n^{*}$ pairs of projected predictor and outcome images by $\{(\bfx^*_i,\bfy_i^{*})\}_{i=1}^{n^{*}}$. Then, we estimate \textbf{IM} by
\begin{align}\label{eq:basis_importance_approx}
    \widehat{\textbf{IM}} = \frac{1}{n^{*}} \sum_{i=1}^{n^{*}} \left| \widehat\bfq(\bfx^{*}_i, \bfy_i^{*}) \right|, \quad \widehat\bfq(\bfx^{*}_i, \bfy_i^{*}) = \frac{1}{S}\sum_{s=1}^{S} \frac{\partial}{\partial \bfx}\log p(\bfy_i^{*} \mid \bfx, \bftheta^{(s)}_{xy}, \bfLambda^{(s)}_{y}) \Big|_{\bfx = \bfx_i^{*}}.
\end{align}
Of note, the function $\frac{\partial}{\partial \bfx}\log p(\bfy \mid \bfx, \bftheta_{xy}, \bfLambda_{y})$ can be efficiently evaluated by using the automatic differentiation algorithm for the deep neural network model. In addition, $(\bfx^*_i, \bfy^*_i)$ may represent new data different from the training data $\mathcal{D}$. When the new data is not available, $(\bfx_i^*,\bfy_i^*)$ can be drawn from $\mathcal{D}$. 

\subsection{Equivalent model representation}\label{sec:method_equivalent}
Combining~\eqref{eq:predictor_img},~\eqref{eq:outcome_img} and~\eqref{eq:dnn_link}, we obtain an equivalent model representation for a better interpretation of the predictor and outcome images in the original space. It is straightforward to show that the predictor images are realizations of GP from (\ref{eq:predictor_img}),
\begin{align}
X_i(v) \sim \GP\{0,\kappa_{x,i}(v,v')\}, \label{eq:X_GP}
\end{align}
where the mean function is zero and the kernel function $\kappa_{x,i}(v,v') = \sum_{k=1}^{K_x}\lambda_{x,k}\psi_{x,k}(v)$ $\psi_{x,k}(v') + \sigma^2_{x,i}I(v = v')$ for any $v, v'\in \calR_x$. To represent the conditional distribution of $Y_i(v)$ given $X_i(v)$, we introduce random effects $\bfe_i = - \sum_{v\in\calR_x} \epsilon_{x,i}(v) \bfpsi_x(v)$. By the property of $\bfpsi_x(v)$ and distribution of $\epsilon_{x,i}$, we have that
\begin{align}
\bfx_i = \sum_{v \in \calR_x} \bfpsi_x(v) X_i(v) + \bfe_i, \quad \bfe_i \sim \mN(\bfzero,\sigma^2_{x,i}\bfI_{K_x}). 
\end{align}
This further implies that the conditional expectation of $Y_i(u)$ given $ \{X_i(v)\}_{v\in \calR_x}$, denoted as $\mu_i(u)$, can be constructed by integrating out $\bfe_i$ in the model. In particular, 
\begin{align}
    \mu_i(u) = \mathbb{E}_{\bfe_i}\left[\scrN\Big\{\sum_{v \in \calR_x} \bfpsi_{x}(v) X_i(v) + \bfe_i;\bftheta_{xy}\Big\}\right]^{\top}\bfpsi_y(u),
\end{align}
where the expectation is taken with respect to $\bfe_i$. Furthermore, by the independence of $\epsilon_{x,i}(v)$ and $\epsilon_{y,i}(u)$, we have
\begin{align}
[ Y_i (u) \mid \{X_i(v)\}_{v\in \calR_x} ] \sim \GP\{ \mu_i(u), \kappa_{y,i}(u,u')\}, \label{eq:Y_GP}
\end{align}
where kernel $\kappa_{y,i}(u,u') = \sum_{k=1}^{K_y}\lambda_{y,k}\psi_{y,k}(u)\psi_{y,k}(u') + \sigma^2_{y,i}I(u = u')$ for any $u, u' \in \calR_y$. We provide the derivations of the equivalent model representation in Supplementary Materials Section S1. 

\subsection{Kernel learning via DNN}\label{sec:metheod_kernel}
The accuracy of approximating images using GPs largely depends on the flexibility of the covariance kernel. Under our framework, it is straightforward to use fixed kernels such as the squared-exponential (SE) and the Mat{\'e}rn kernel. However, such kernels lack the flexibility to explain complex spatial structures in practice. Therefore, we develop a novel DNN-based data adaptive method to estimate the eigenfunctions and construct the covariance kernel of GPs in \eqref{eq:X_GP} and \eqref{eq:Y_GP}. We only illustrate our method on the construction of $\bfpsi_x(v)$ for predictor images. A similar approach can be applied to outcome images. For $v\in \calR_x$, let $\bfX(v) = \{X_1(v),\ldots, X_n(v)\}^\top$ be a vector of $n$ observed predictor image measurements at voxel $v$. To approximate $\bfX(v)$, we introduce a feed-forward DNN $\scrN(v;\bftheta_x): \mbR^{d_x} \mapsto \mbR^{K_x}$, where the dimension of the input layer is the dimension of the predictor image voxel $d_x$ and the dimension of the output layer is the number of orthonormal basis functions $K_x$. Note that the DNN here for kernel learning is different from the one we use in (\ref{eq:dnn_link}). Then we adopt the linear transformation to project $\scrN(v;\bftheta_x)$ onto $\mbR^n$ for approximating $\bfX(v)$. We solve the following optimization problem,
\begin{align}\label{eq:DNN_approx}
(\hat \bfP_x, \hat\bftheta_x) = \underset{(\bfP_x, \bftheta_x)}{\arg\min} \sum_{v\in\calR_x}\big\|\bfX(v) - \bfP_x \scrN(v;\bftheta_x)\big\|^2_2, 
\end{align}
where $\bfP_x \in \mbR^{n\times K_x}$ is the linear projection matrix. The estimated DNN function $\scrN(v;\hat\bftheta_x)$ can be considered as a vector of $K_x$ unorthonormalized basis functions. To construct $\bfpsi_x(v)$, for general $\calR_x$, we can apply the Gram-Schmidt process on $\scrN(v;\hat\bftheta_x)$ for orthonormalization. When $\calR_x$ contains a finite number of equal space grid points, we use the Singular Value Decomposition on matrix $\{\scrN(v;\hat\bftheta_x)\}_{v\in\calR_x}$ to obtain the orthonormal matrix $\{\bfpsi_x(v)\}_{v\in\calR_x}$. 

Note that the classical Principle Component Analysis (PCA) can decompose the observed images and provide a set of orthonormal basis functions but is subject to noise and is prone to overfit. However, constructing the kernel via DNNs, our approach implicitly imposes smoothness constraints and is more robust. Furthermore, we may add regularization terms to the objective function or apply dropout layers [\cite{srivastava2014dropout}] to prevent overfitting. Another advantage of our approach over PCA is that we learn the kernel as a function so that it is possible to interpolate or extrapolate kernel function values, while PCA only provides the value of the kernel function evaluated at fixed locations. 

\subsection{Prior specifications}\label{sec:method_prior}
Given the estimated orthonormal basis functions for predictor images and outcomes $\bfpsi_x(v)$ and $\bfpsi_y(u)$, we perform Bayesian inferences on the proposed model. For the weight and bias parameters of the DNN model in (\ref{eq:dnn_link}), we assign independent normal priors, i.e., for $\ell = 1,\ldots, L$
%\begin{align*}
 $\mvec(\bfW_{xy,\ell}) \sim \mN(\bfzero, \sigma^2_w\bfI_{S_{xy,\ell}\times S_{xy,\ell-1}})$,$\bfb_{xy,\ell} \sim \mN(\bfzero, \sigma^2_w\bfI_{S_{xy,\ell}}),$
%\end{align*}
where $\sigma^2_w$ is the prior variance parameters for the weight and bias parameters. For all the variance parameters in the model, we assign inverse gamma priors, i.e., for $i = 1,\ldots, n$, $k_x = 1,\ldots, K_x$ and $k_y = 1,\ldots, K_y$, we have
%\begin{align*}
$\sigma^2_{x,i}, \sigma^2_{y,i} \sim \IG(a_\sigma, b_\sigma)$, $\sigma^2_w\sim \IG(a_w, b_w)$, $\lambda_{x,k_x}, \lambda_{x,k_y} \sim \IG(a_\lambda, b_\lambda)$, 
%\end{align*}
where all the shape and scale parameters in the inverse gamma distributions are pre-specified. In practice, we suggest to set  $a_\sigma=b_\sigma=a_w=1$. While suitable $a_\lambda$, $b_\lambda$ and $b_w$ may vary across different datasets, we suggest $b_w$ take value from 1 to 100, and $a_\lambda=b_\lambda=1$ or $a_\lambda = 1$ and $b = 0.1$. We provide sensitivity analysis in Supplementary Materials S2, showing that mild changes in hyperparameters will not lead to a drastic decrease in the model performance. 

\section{Posterior Computation}\label{sec:posterior}
To ensure the efficiency and scalability of BIRD-GP, we develop a two-stage hybrid posterior computation algorithm. The two-stage algorithm is efficient in terms of both time and memory. The computational bottleneck resides in the projection of high-dimensional images onto the low-dimensional Euclidean space, which can be greatly mitigated by Stage 1 posterior computation that can be straightforwardly paralleled across subjects. In Stage 1, we apply the Gibbs sampler for Bayesian linear regressions (\ref{eq:predictor_img}) and (\ref{eq:outcome_img}) to simulate the posterior distribution of the projected predictor image $\bfx_i$ and outcome image $\bfy_i$ along with associated variance parameters $\bfLambda_x$, $\bfLambda_y$, $\sigma^2_{x,i}$ and $\sigma^2_{y,i}$ given all other parameters. 

\begin{algorithm}[t]
\small
\spacingset{1}
\begin{algorithmic}
    \STATE Draw $S$ random samples from the prior distribution: $\bfTheta^0_1,\ldots, \bfTheta^0_S$. 
    \STATE Update $\{\bfTheta^0_s\}_{s=1}^S$ for $T$ iterations
    \FOR{$t$ in $0:(T-1)$}
        \STATE Sample indices $\cI \subset [n]=\{1,\ldots,n\}$
        \FOR{$s$ in $1:S$}
          \STATE Compute $\bfTheta^{t+1}_s$ given $\bfTheta^{t}_1,\ldots, \bfTheta^{t}_S$ and the subset of projected images $\bfD_{\cI}$ via~\eqref{eq:update_theta}
        \ENDFOR
    \ENDFOR
\end{algorithmic}
\caption{SVGD for Bayesian neural networks}
\label{svgdbnn}
\end{algorithm}

In Stage 2, we adopt the Stein Variational Gradient Descent Algorithm (SVGD) [\cite{liu2016stein}] to simulate the posterior distribution of all the parameters that are associated with the DNN model in (\ref{eq:dnn_link}), i.e., $\bfTheta=\{\bftheta_{xy},\sigma^2_w,\sigma^2_b\}$. As a gradient-based method, SVGD provides a general variational inference framework. Let $\bfD_{\cI} = \big\{\mathbb{E}[\bfx_i\,|\,X_i(v)],\,\mathbb{E}[\bfy_i\,|$ $Y_i(u)]\big\}_{i\in \cI}$ represent a collection of posterior mean of the projected predictor images and the projected outcome images with indices in $\cI\subset [n] =\{1,\ldots, n\}$. Denote by $q(\bfTheta)$ and $p(\bfTheta\,|\,\bfD_\cI)$ the prior distribution and the posterior distribution of $\bfTheta$ given data $\bfD_{\cI}$, respectively. The first step is to draw $S$ random samples from the prior $q(\bfTheta)$, denoted as $\{\bfTheta_s^0\}_{s=1}^S$. Then we iteratively update the samples of $\bfTheta$ from the prior towards the posterior distribution using the stochastic gradient descent (SGD) algorithm. Specifically, at the $(t+1)$th iteration $(t = 0,\ldots, T-1)$, we first sample $\cI \subset [n]$; then for $s = 1,\ldots, S$, we update the $s$th sample of $\bfTheta$, denoted by $\bfTheta^{t}_{s}$, by the following rule, 
\begin{equation}\label{eq:update_theta}
    \bfTheta^{t+1}_{s} = \bfTheta^{t}_{s} + \frac{\alpha}{S}\sum^{S}_{s'=1} \bigg[k(\bfTheta^{t}_{s'}, \bfTheta^{t}_{s}) \big\{\nabla_{\bfTheta^{t}_{s'}}\log p(\bfTheta^{t}_{s'}\,|\,\bfD_{\cI})\big\} +\big\{\nabla_{\bfTheta^{t}_{s'}} k(\bfTheta^{t}_{s'}, \bfTheta^{t}_{s})\big\}\bigg],
\end{equation}
where $\alpha$ is the step size and $k(\cdot, \cdot)$ is the kernel function that defines the Stein discrepancy between the density of $\{\bfTheta_s^t\}_{s=1}^S$ and the target density $p(\bfTheta\,|\,\mathcal{D}_{[n]})$. Here, we present the basic SGD algorithm in~\eqref{eq:update_theta}. In practice, one may choose more advanced SGD algorithm such as the Adam optimizer [\cite{kingma:adam}]. We use the Gaussian kernel $k(x, y) = \exp\big\{-\frac{1}{h}\|x-y\|^2_2\big\}$ where $h$ is chosen as the median of the pairwise distance between the current samples $\{\bfTheta^t_s\}_{s=1}^S$ divided by $\log(n)$ [\cite{liu2016stein}]; the closed-form of $\nabla_{\bfTheta^{t}_{s'}} k(\bfTheta^{t}_{s'}, \bfTheta^{t}_{s})$ is available in this case. We adopt the automatic differentiation approach to compute the gradient of $\log p(\bfTheta^{t}_{j}\,|\,\bfD_{\cI})$ in practice. 

\section{Simulations on Synthetic Data}\label{sec:synthetic}
We evaluate the performance of BIRD-GP on synthetic data based on MNIST and Fashion MNIST datasets, and compare BIRD-GP with three DNNs, three CNNs, and the Recursively Branched Deconvolutional Network (RBDN) [\cite{santhanam2017generalized}]. DNNs and CNNs are with different architectures. Under the BIRD-GP framework, we also compare the DNN-based kernel and other kernel constructions -- the squared-exponential (SE) and the Mat{\'e}rn kernel, and the PCA-based kernel. The hyperparameters of all competitors are discussed the Supplementary Materials S3.

\begin{figure}[t]
    \small
    \spacingset{1}
    \centering
    \begin{subfigure}{\textwidth}
        \centering
        \includegraphics[scale = 0.3]{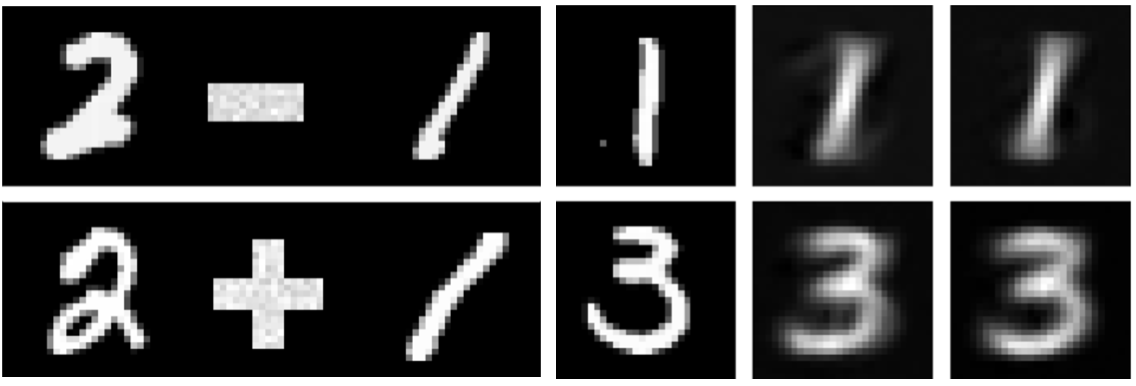}
        \caption{Two test examples from the MNIST experiment. The first two columns show the synthesized predictor images and outcome images. The third column shows the predicted outcomes by BIRD-GP, while the last column is the predicted outcomes by CNN2.}
        \label{fig:mnist}
    \end{subfigure}
    \vfill
    \vspace{0.2cm}
    \begin{subfigure}{\textwidth}
        \centering
        \includegraphics[scale = 0.25]{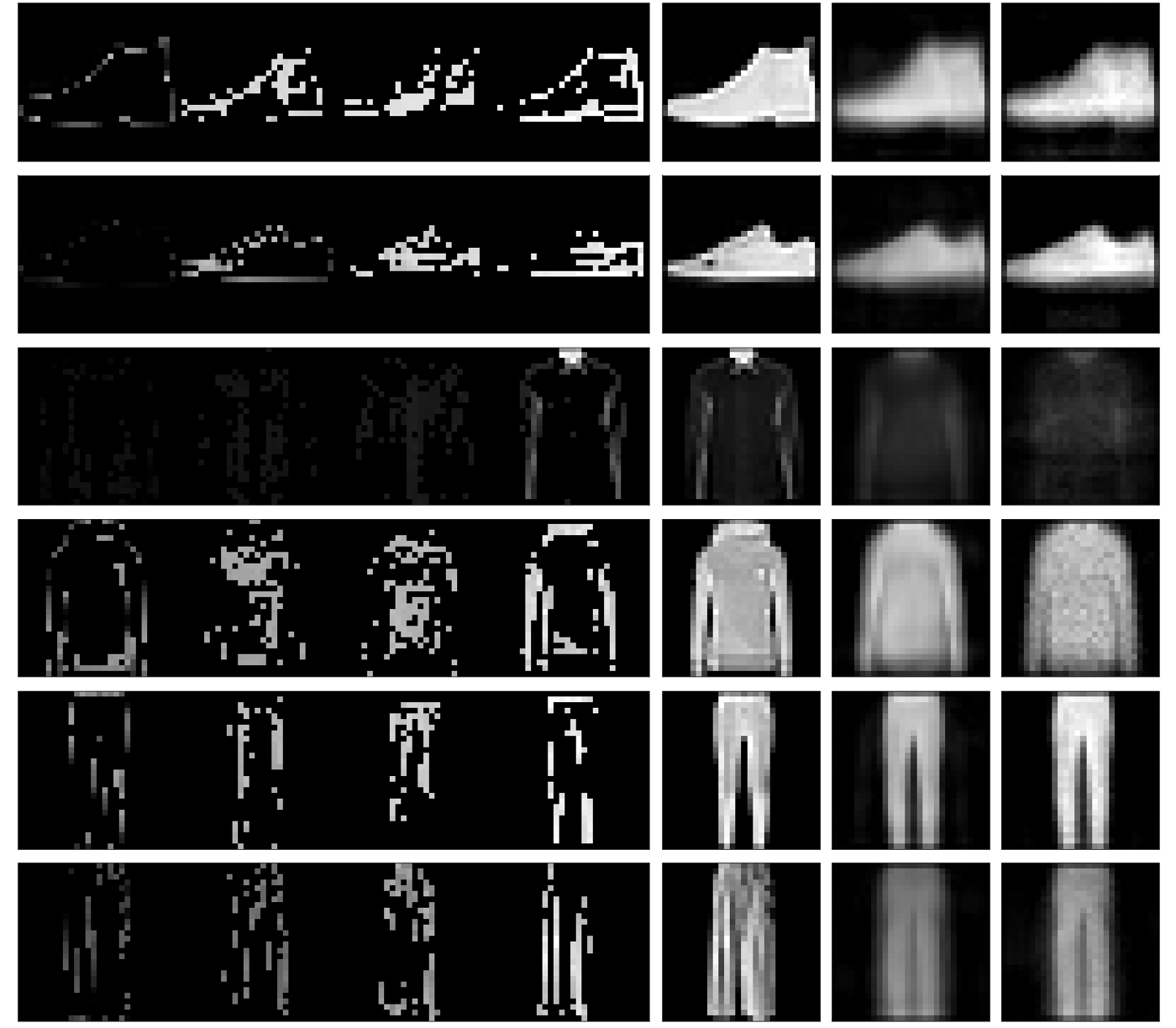}
        \caption{Six test examples from the Fashion MNIST experiment. The first two columns show the synthesized predictor images and outcome images. The third column shows the predicted outcomes by BIRD-GP, while the last column is the predicted outcomes by CNN2.}
        \label{fig:fmnist}
    \end{subfigure}
    \caption{Example images from the (a) MNIST experiment and (b) Fashion MNIST experiment.}
\end{figure}

\begin{figure}[t]
    \small
    \spacingset{1}
    \centering
    \begin{subfigure}{0.6\textwidth}
        \centering
        \includegraphics[scale = 0.35]{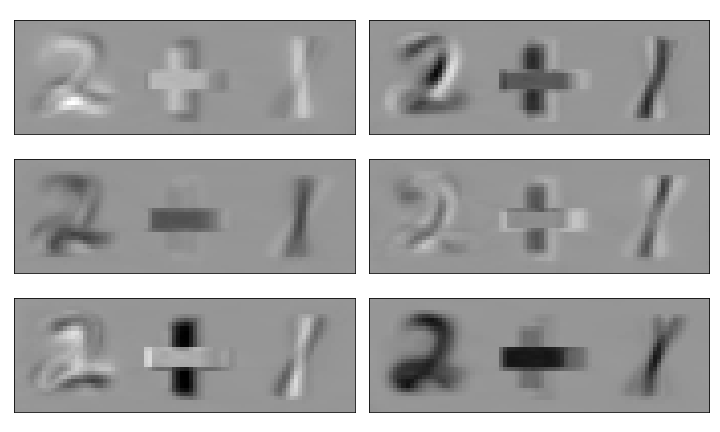}
        \caption{}
        \label{fig:basis_mnist_in}
    \end{subfigure}
    \hfill
    \begin{subfigure}{0.38\textwidth}
        \centering
        \includegraphics[scale = 0.35]{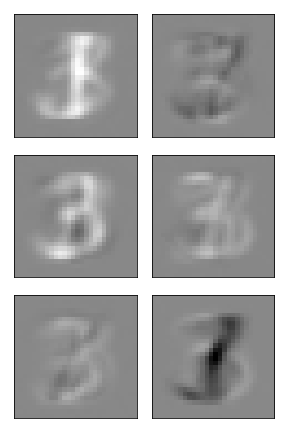}
        \caption{}
        \label{fig:basis_mnist_out}
    \end{subfigure}
    \caption{The first 6 basis images with the highest eigenvalues in the MNIST experiment for (a) predictors and (b) outcomes.}
    \label{fig:basis_mnist}
\end{figure}

\subsection{MNIST} 
The MNIST handwritten digits dataset [\cite{lecun2010mnist}] contains 60,000 training and 10,000 testing image-label pairs. Each image is a $28 \times 28$ handwritten digit (0-9). We design the experiment so as to mimic the calculation of ``$2+1=3$" and ``$2-1=1$". We synthesize predictor images of size $28 \times 84$ based on the MNIST dataset by horizontally stacking an image of ``2", an $28 \times 28$ image of either ``$+$" or ``$-$" and an image of ``1". The plus/minus sign in between are randomly generated with varying margins, widths and lengths. The outcome images are either images of ``3" or ``1", depending the sign in the middle of the predictor images. Images of ``1", ``2" and ``3" in predictors and outcomes are randomly selected without replacement. Figure \ref{fig:mnist} shows two examples in the testing set. 

We generate 1,000 training samples and 1,000 testing samples for each dataset and repeat the experiment on 50 datasets. We use the fully trained models to predict the outcome images; the predictions are further fed into a pre-trained CNN binary classifier of MNIST images of ``1" and ``3". The pre-trained CNN classifier uses the outcome images for training and testing, and obtains close-to-perfect training and testing accuracy. We evaluate the performance of all models by checking the classification accuracy of the predicted label by the pre-trained classifier given the predicted outcome images, as this is a good indicator of the extent to which the model restores the outcome images. 

\begin{table}[t!]
    \small
    \spacingset{1}
    \centering
    \caption{Training and testing performance of BIRD-GP and other competing methods are summarized for (a) MNIST and (b) Fashion MNIST experiments.}
    \begin{subtable}{0.417\textwidth}
        \centering
        \caption{Proportions of replicates where classification accuracy is larger than 0.995 for both training and testing synthesized MNIST images. Number of parameters are shown after the model name.}
        \resizebox{\textwidth}{!}{%
        \begin{tabular}{l c c }
        \hline\hline
                              & \textbf{Training} & \textbf{Testing}  \\
        \hline
        \textbf{BIRD-GP (126K)} & \textbf{1.00} & \textbf{1.00} \\
        \textbf{PCA-BIRD-GP}  & \textbf{1.00}   & 0.02          \\
        \textbf{SE-BIRD-GP}   & 0.96            & 0.96          \\
        \textbf{Mat{\'e}rn-BIRD-GP} & \textbf{1.00} & \textbf{1.00}      \\
        \textbf{DNN1 (202K)}  & 0.02             & 0.00             \\
        \textbf{DNN2 (935K)}  & 0.02             & 0.00             \\
        \textbf{DNN3 (1001K)} & 0.02             & 0.08             \\
        \textbf{CNN1 (168K)}  & 0.92          & 0.88          \\
        \textbf{CNN2 (99K)}   & 0.60          & 0.80          \\
        \textbf{CNN3 (225K)}  & 0.88          & 0.96          \\
        \textbf{RBDN (445K)}  & \textbf{1.00}    & \textbf{1.00}    \\
        \hline
        \bottomrule
        \end{tabular}%
        }
        \label{mnist_table}
    \end{subtable}
    \hfill
    \begin{subtable}{0.57\textwidth}
        \centering
        \caption{Mean and standard deviation of training and testing MSE ($\times10^{-4}$) over 50 synthesized datasets based on Fashion MNIST. Number of parameters are shown after the model name.}
        \resizebox{\textwidth}{!}{%
        \begin{tabular}{l c c }
        \hline\hline
        & \textbf{Training MSE}    & \textbf{Testing MSE}     \\
        \hline
        \textbf{BIRD-GP (126K)}  & 234 (10)       & \textbf{260 (10)} \\
        \textbf{PCA-BIRD-GP}     & \textbf{203 (38)} & 301 (39)       \\
        \textbf{SE-BIRD-GP}      & 356 (11)       & 379 (9)          \\
        \textbf{Mat{\'e}rn-BIRD-GP} & 354 (10)    & 378 (10)          \\
        \textbf{DNN1 (252K)}  & 341 (18)          & 367 (14)          \\
        \textbf{DNN2 (1136K)} & 243 (5)          & 306 (5)          \\
        \textbf{DNN3 (1202K)} & 238 (4)          & 310 (5)          \\
        \textbf{CNN1 (193K)}  & 348 (16)          & 364 (15)          \\
        \textbf{CNN2 (103K)}  & 324 (9)          & 343 (7)          \\
        \textbf{CNN3 (233K)}  & 267 (8)          & 300 (7)          \\
        \textbf{RBDN (445K)}  & 307 (41)          & 475 (53)          \\
        \hline
        \bottomrule
        \end{tabular}%
        }
        \label{fashion_mnist_table}
    \end{subtable}
    \vfill
    \vspace{0.3cm}
    \begin{subtable}{\textwidth}
        \centering
        \caption{Fashion MNIST experiment MSE ($\times10^{-4}$) stratified by image label class (train / test), sorted by testing MSE in descending order.}
        %\resizebox{\columnwidth}{!}{%
        \begin{tabular}{r c c c c c}
        \hline\hline
        \textbf{Label}      & Bag              & Ankle Boot        & Sandal            & Dress             & Coat              \\
        \hline
        \textbf{MSE}        & 339 / 420  & 314 / 347   & 322 / 345   & 244 / 278   & 233 / 277   \\
        \hline\hline
        \textbf{Label}      & Pullover         & T-shirt           & Shirt             & Trouser           & Sneaker           \\
        \hline
        \textbf{MSE}        & 200 / 252  & 218 / 248   & 214 / 230   & 196 / 226   & 222 / 222   \\
        \hline
        \bottomrule
    \end{tabular}%
    %}
    \label{mse_by_label}
    \end{subtable}
\end{table}

We use a four-layer neural network with ReLU activation for kernel learning. Each layer has 128 hidden nodes. The number of eigenfunctions is set to be 50 for both predictors and outcomes. For the Bayesian neural network, we adopt a ReLU-activated one-layer structure with 200 neurons. We train the BNN for 30 epochs with batch size 64 by SVGD. The six neural networks are trained for 100 epochs and RBDN is trained for 50 epochs, all with batch size 64. In Table \ref{mnist_table}, we summarize the proportion of experiments where the classification accuracy is larger than 0.995 over the 50 replicates. 
Both the DNN-based kernel and the Mat{\'e}rn kernel under the BIRD-GP framework achieve training and testing accuracy of at least 0.995 for all datasets. Even with a relatively small number of parameters, BIRD-GP outperforms all CNNs and DNNs. RBDN has similar performance to BIRD-GP, but with much more paramters. The PCA-based kernel performs well on the training set, but has poor performance on the testing set. Figure \ref{fig:basis_mnist} shows the top six basis images with the largest eigenvalues for predictors and outcomes, respectively. The predictor basis images show shapes of ``2'' on the left, ``+'' or ``-'' in the middle and ``1'' on the right. The outcome basis images show shapes of ``1'' and ``3'' in the middle. The basis images remain near to constant in other areas where there are little variability in the original predictor and outcome images.

\begin{figure}[t!]
    \small
    \spacingset{1}
    \centering
    \begin{subfigure}[t]{0.6\textwidth}
        \centering
        \includegraphics[scale = 0.28]{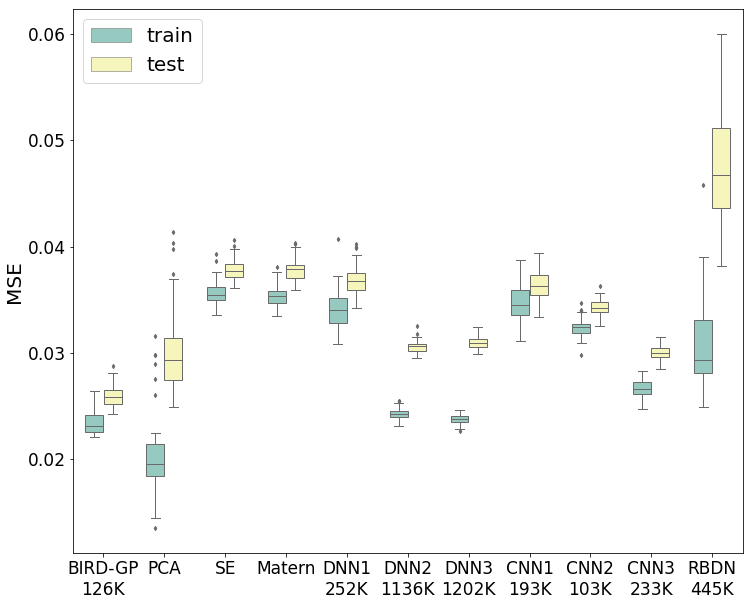}
        \caption{}
        \label{fig:fmnist_mse}
    \end{subfigure}
    \hfill
    \begin{subfigure}[t]{0.38\textwidth}
        \centering
        \includegraphics[scale = 0.4]{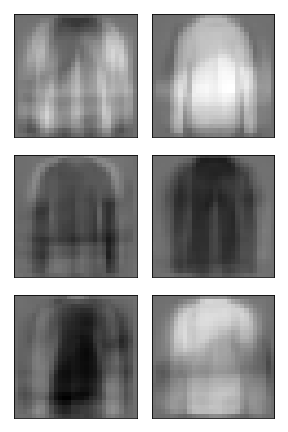}
        \caption{}
        \label{fig:bas_img_out}
    \end{subfigure}
    %\vskip\baselineskip
    %\vspace{0.1pt}%
    \vfill
    \begin{subfigure}{\textwidth}
        \centering
        \includegraphics[scale = 0.52]{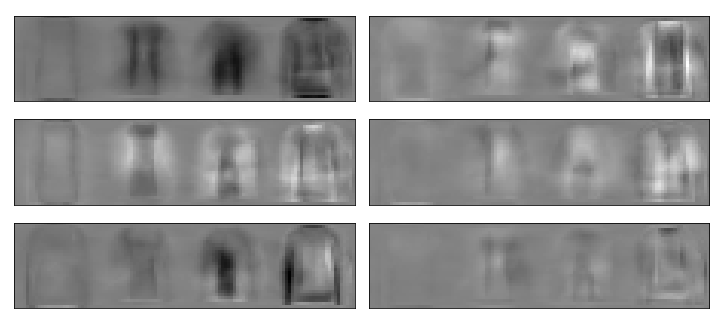}
        \caption{}
        \label{fig:bas_img_pred}
    \end{subfigure}
    \caption{The Fashion MNIST experiment: (a) The boxplot of training and testing MSE over 50 datasets for the Fashion MNIST experiment. We compare BIRD-GP with other methods. Under the BIRD-GP framework, we consider the DNN-based kernel, the PCA-based kernel, the squared-exponential (SE) kernel and the Mat{\'e}rn kernel. DNN1, DNN2 and DNN3 are deep neural network models with different architectures. CNN1, CNN2 and CNN3 are deep convolutional neural network models with different architectures. RBDN stands for the recursively branched deconvolutional network model. The number of parameters for neural network models are presented below model names. (b) The first 6 outcome basis images with the highest eigenvalues. (c) The first 6 predictor basis images with the highest eigenvalues. }
\end{figure}

\subsection{Fashion MNIST} 
The Fashion MNIST dataset [\cite{xiao2017fashion}] is a gray-scale image dataset containing 10 classes of fashion products. The dataset contains 60,000 training images and 10,000 testing images, each of size $28 \times 28$. We synthesize 1,000 training samples and 1,000 testing samples based on the dataset. For each image, we filter four sub-images based on quartiles of non-zero voxel intensities. Each sub-image keeps non-zero voxel intensities within an inter-quartile range, while other voxels are masked to 0. We stack the four sub-images horizontally, with the first inter-quartile image on the left and the fourth inter-quartile image on the right. We treat the stacked image with size $28 \times 112$ as the predictor while the original image as the outcome. Figure \ref{fig:fmnist} shows six examples from the testing set. 

We generate 50 such datasets and compare the MSE on the predictions of outcome images by BIRD-GP and other methods. Training parameters of all methods are the same as Section 4.1. Table \ref{fashion_mnist_table} shows the mean and standard deviation of MSE over 50 replicates. Figure \ref{fig:fmnist_mse} shows the boxplot for the training and testing accuracy over 50 replicates. BIRD-GP performs the best on the testing data despite its limiting number of parameters. Figure \ref{fig:fmnist} shows BIRD-GP can well capture the individual difference of test outcome images, even with limited training samples. It is worth noting that the PCA-based kernel performs the best on the training data while it is not competitive on the test set, and fixed kernels lack the flexibility to fit the data well. 

To demonstrate 50 basis functions are sufficient, we repeat the experiment on the first dataset using 100 basis functions. The first 50 basis account for 95.66\% of total variance in the predictor images and 98.37\% in the outcome images (see Supplementary Materials S4). Figures \ref{fig:bas_img_pred} and \ref{fig:bas_img_out} show the top six basis images with the largest eigenvalues for predictors and outcomes, respectively. We see patterns of shirts, trousers and sneakers in the top basis images. In Table \ref{mse_by_label}, we summarize within-class MSEs. BIRD-GP performs better on images whose patterns are detected by the top eigenfunctions, but loses some of its power in other types of images. Finally, to demonstrate our method can measure prediction uncertainty, we compute the mean coverage rates (MCR) of the voxel-wise 95\% predictive credible interval for each outcome image. The average of MCR across images is 97.7\% (s.d. 1.1\%) on the training set and 95.9\% (s.d. 3.2\%) on the testing set. 

We record the computing time of BIRD-GP and other methods on a Macbook Pro with the M1 Pro chip and 16GB RAM. For one dataset (1000 training samples), BIRD-GP requires 628 seconds for training. Specifically, kernel learning for predictors takes 151 seconds, kernel learning for outcomes takes 64 seconds, refitting predictor basis coefficients after orthogonalization takes 192 seconds, refitting outcome basis coefficients after orthogonalization takes 193 seconds, and SVGD requires 24 seconds. In comparison, RBDN takes 737 seconds to train, while the training times for DNN1, DNN2, and DNN3 are 3 seconds, 7 seconds, and 16 seconds, respectively. CNN1, CNN2, and CNN3 require 234 seconds, 211 seconds, and 385 seconds. The computing time of BIRD-GP is comparable to that of RBDN and CNN. We discuss potential improvement on computing time in Section~\ref{sec:conclusion}.

\subsection{HCP data based simulations}\label{sec:synthetic_hcp}

To evaluate the performance of BIRD-GP on neuroimaging data, we generate synthetic images based on the fMRI data in the Human Connectome Project (HCP) analyzed in Section~\ref{sec:hcp}. We consider three scenarios. In Scenarios 1 and 2, we simulate data using BIRD-GP, i.e., models~\eqref{eq:predictor_img} -- \eqref{eq:dnn_link}. In Scenario 3, we simulate data using voxel-wise regression model.  We set the sample size $n=714$, the number of basis functions for predictors $K_x = 150$, and the number of basis functions for outcomes $K_y = 150$ which are consistent with the settings in real data analysis in Section~\ref{sec:hcp}. Training details and hyperparameters are the same with those in Section~\ref{sec:hcp}. In all three scenarios, we consider the resting state fALFF as predictors and the math-story contrast map in the language task as outcomes. Both the predictor images and outcome images are of dimension $91\times109\times91$. Details on these images are described in Section \ref{sec:hcp_data}. In this section, we compare BIRD-GP with two commonly used approach in neuroimaging comunity: linear regression (LR) [\cite{tavor2016task}] and voxel-wise regression (VR) [\cite{dworkin2016prevail}].

In Scenarios 1 and 2, we generate the projected predictor images $\bfx_i\sim\mN(\bfzero_{K_x}, \bfI_{K_x})$ for $i=1,\ldots,n$. Then, we simulate the projected outcome images $\bfy_i\mid\bfx_i\sim\mN(\bfB\bfx_i, \bfLambda_{K_y})$ where $\bfB$ specifies a linear mapping from $\bfx_i$ to $\bfy_i$ and $\bfLambda_{K_y}=\bfI_{K_y}$ for Scenarios 1, and $\bfy_i\mid\bfx_i\sim\mN\{\scrN(\bfx_i;\hat{\bftheta}_{xy}), \bfLambda_y\}$ where the fully-trained neural network $\scrN(\bfx_i;\hat{\bftheta}_{xy})$ obtained from Section \ref{sec:hcp_fit} specifies the association and $\bfLambda_{y}$ is determined such that the signal-to-noise ratio is 0.5. Finally, the predictors images $X_i(v)=\bfx_i^\top\hat{\bfpsi}_x(v)$ and the outcome images $Y_i(v)=\bfy_i^\top\hat{\bfpsi}_y(v)$, where $\hat{\bfpsi}_x(v)$ is the fitted basis functions for fALFF and $\hat{\bfpsi}_y(v)$ is the fitted basis functions for the story-math contrast, both from Section \ref{sec:hcp_fit}.

In Scenario 3, we simulate outcome image $Y_i(v)$ based on a voxel-wise regression model: 
$$Y_i(v) = \beta_0(v) + \beta_1(v)X_i(v) + r_i(v),$$ where $X_i(v)$ represents the fALFF value at voxel $v$ from subject $i$, $\beta_0(v)$ and $\beta_1(v)$ are independently generated from the standard Gaussian distribution for all $v$ ; and random noises $r_i(v)$ follow the standard Gaussian distribution for all $i$ and $v$.

\begin{figure}[!ht]
    \small
    \spacingset{1}
    \centering
    \begin{subfigure}[t]{\textwidth}
        \centering
        \includegraphics[width=0.9\linewidth]{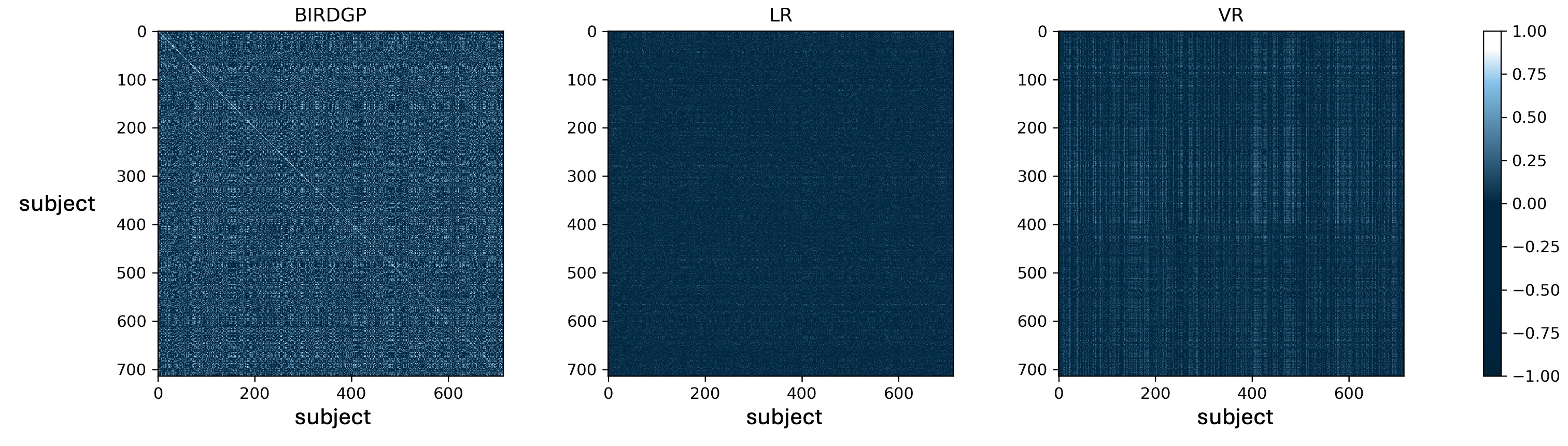}
        \caption{Scenario 1.}
        \label{fig:correlation_simulation2}
    \end{subfigure}
    \vfill
    \begin{subfigure}[t]{\textwidth}
        \centering
        \includegraphics[width=0.9\linewidth]{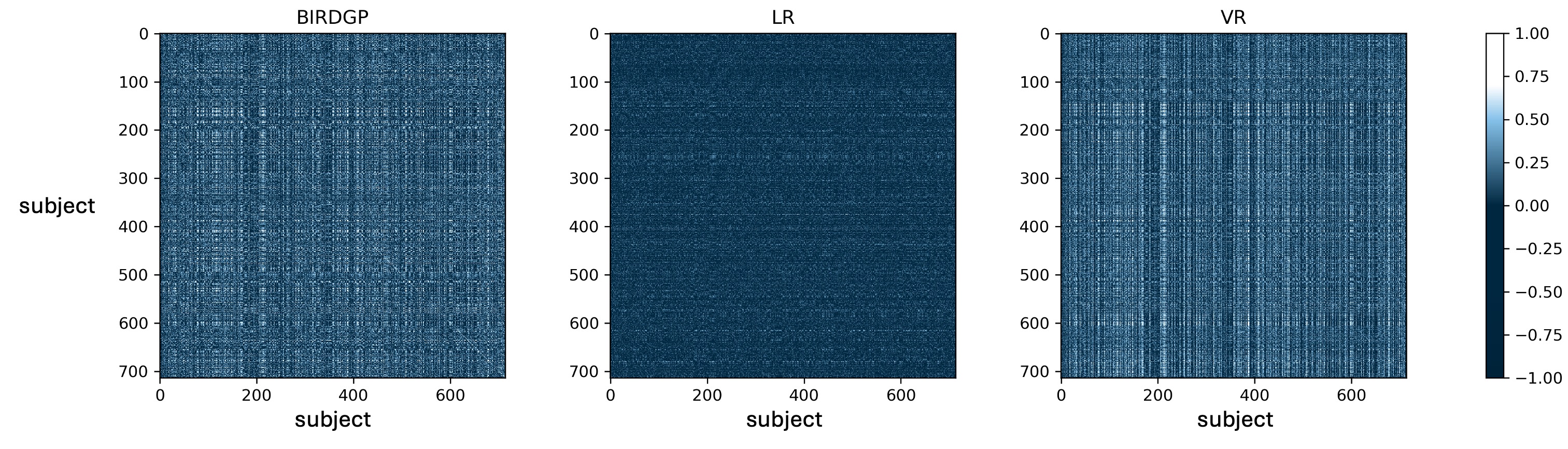}
        \caption{Scenario 2.}
        \label{fig:correlation_simulation3}
    \end{subfigure}
    \vfill
    \begin{subfigure}[t]{\textwidth}
        \centering
        \includegraphics[width=0.9\linewidth]{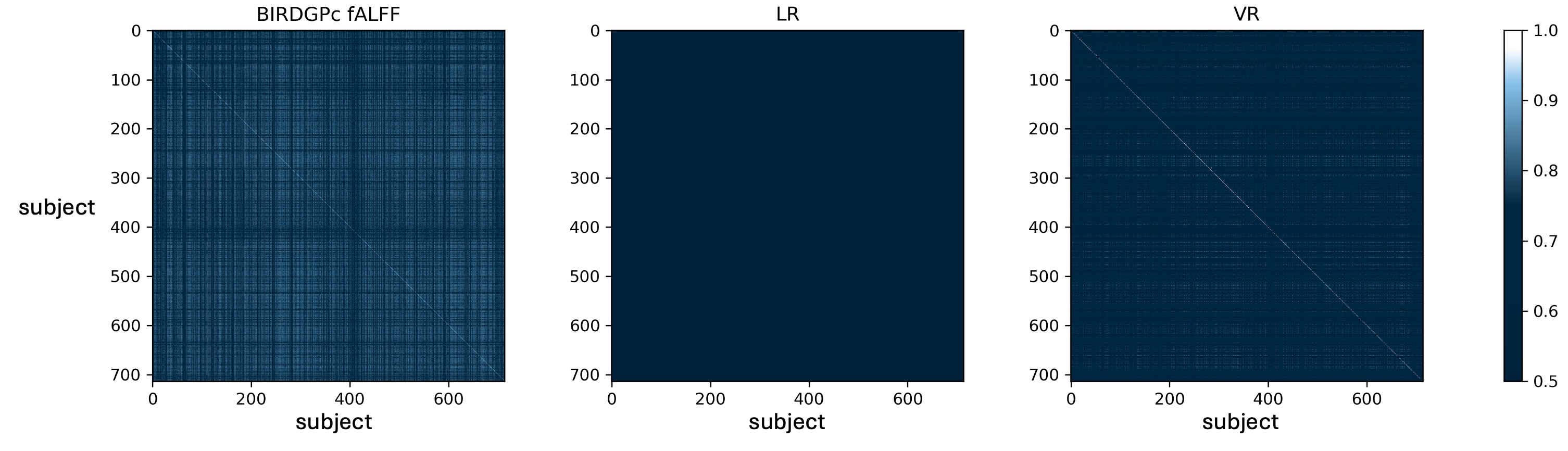}
        \caption{Scenario 3.}
        \label{fig:correlation_simulation4}
    \end{subfigure}
    \caption{The Pearson correlation between the predicted maps (rows) and actual outcome maps (columns) for all pairs of subjects of BIRD-GP and competing methods in three scenarios: (a) Scenario 1, (b) Scenario 2, and (c) Scenario 3. Each entry in the matrix is the correlation between the outcome map of a subject and the same subject (diagonal), or the predicted map of another (off-diagonal). The diagonal-dominant pattern for BIRD-GP indicates BIRD-GP's prediction for any subject is more similar to the subject’s own outcome map than to other subjects’ outcome maps.}
    \label{fig:correlation_simulation}
\end{figure}

To quantitatively assess the predictive performance of BIRD-GP, we compute the Pearson correlation between the predicted and observed outcome maps for all pairs of subjects. We partition the subjects into five folds. Each time, we use one fold of data as test data and the rest as training data. We define a prediction correlation matrix $\bfC = (c_{ij})$ between the predictions and observed maps for image-on-image regressions, where $c_{ij}$ is the Pearson correlation between the predicted outcome map of subject $i$, based on their predictor image, and the observed outcome map of subject $j$. This definition has been adopted by \citet{tavor2016task}. The diagonal entry $c_{ii}$ represents the correlation between the outcome map and the predicted map of subject $i$ while off-diagonal elements $c_{ij}$ for $i\neq j$ are the correlation between the predicted map of subject $i$ and the observed map of other subjects. A model fitting resulting with $c_{ii} > c_{ij}$ for most $j\neq i$ if not all indicates a strong prediction performance. We define the subject-specific predictable activated region as the set of voxels that exhibit the top 5\% absolute intensity values for the subject and are also predicted to have top 5\% absolute intensity values by BIRD-GP. Figure \ref{fig:correlation_simulation} compares the heatmaps of the prediction correlation matrix for BIRD-GP and other methods on predictable activated regions. 

In all scenarios, the correlation matrix of BIRD-GP is diagonal-dominant, indicating that BIRD-GP's predictions for each subject are more similar to the subject’s own outcome map than to the outcome maps of other subjects. In Scenarios 1 and 2, the diagonal elements in the correlation matrix of BIRD-GP are brighter than those of LR and VR, showing that BIRD-GP outperforms other methods in prediction. In Scenario 3 where the data is generated by a mis-specified model, BIRD-GP shows comparable performance to the true model VR. This shows BIRD-GP is robust under model mis-specification.

\begin{figure}[t]
    \centering
    \small
    \spacingset{1}
    \includegraphics[width=\linewidth]{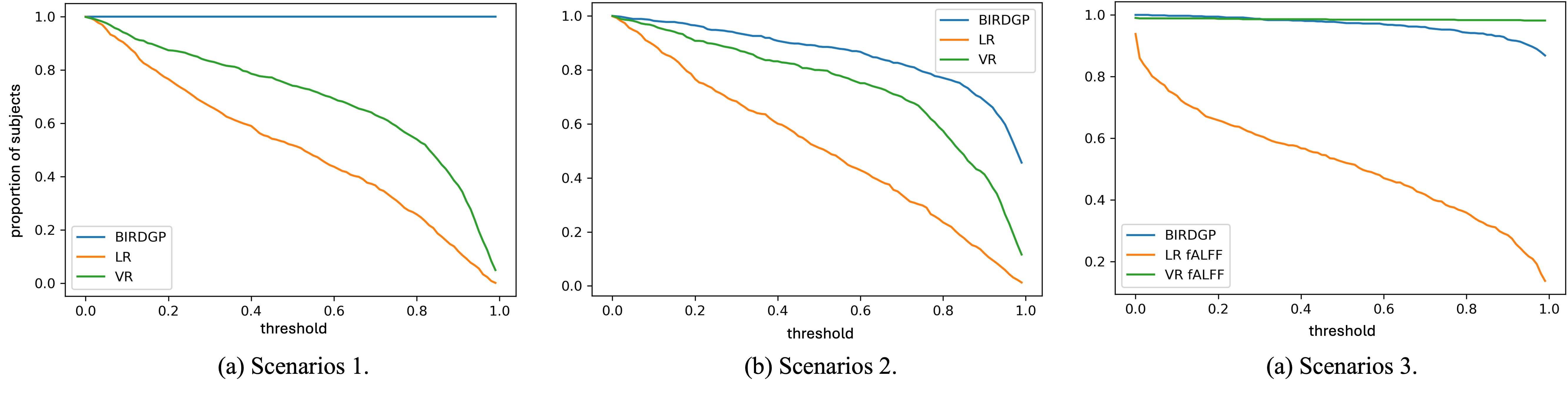}
%    \begin{subfigure}{0.32\textwidth}
%        \centering
%        \includegraphics[width=\linewidth]{figures/proportion_simulation2_label.jpg}
%        \caption{Scenario 1.}
%    \end{subfigure}
%    \hfill
%    \begin{subfigure}{0.32\textwidth}
%        \centering
%        \includegraphics[width=\linewidth]{figures/proportion_simulation3_label.jpg}
%        \caption{Scenario 2.}
%    \end{subfigure}
%    \hfill
%    \begin{subfigure}{0.32\textwidth}
%        \centering
%        \includegraphics[width=\linewidth]{figures/proportion_csimulation4_label.jpg}
%        \caption{Scenario 3.}
%    \end{subfigure}
    \caption{The proportion of subjects $p(\alpha)$ (y-axis) that have a larger ``diagonal'' correlation than ``off-diagonal'' correlations in the same row, over different thresholds $\alpha$ (x-axis) ranging from 0-1. The ``diagonal'' correlation represents the Pearson correlation between the outcome map of a subject and the predicted map of same subject, while the ``off-diagonal'' elements are the correlation between the predicted map of a subject and the observed map of others. The proportion measures the chance of obtaining a better prediction on the outcome map using a model prediction than using a random outcome map from other subjects. }
    \label{fig:proportion_simulation}
\end{figure}

To further evaluate the predictive performance of BIRD-GP, we define a subject-specific accuracy measurement by the frequency that the correlation between the predicted map of a subject and the outcome map of the same subject is larger than the correlation between the predicted map of a subject and the outcome map of others, i.e., the frequency of $c_{ii}$ larger than $c_{ij}$ for $i\neq j$, 
\begin{align}
    a_i = \frac{\sum_{j\neq i}^{n}\text{I}_{\{c_{ii}>c_{ij}\}}}{n-1}.
\end{align}
We then compute the proportion of subjects $p(\alpha)$ that have a larger $a_i$ over different thresholds $\alpha$ from 0-1, i.e.,
\begin{align}
    p(\alpha) = \frac{\sum_{i=1}^n\text{I}_{\{a_i > \alpha\}}}{n}.\label{eq:prop_subjects}
\end{align}
This proportion measures the chance of obtaining a better prediction on the outcome map using a model prediction than using a random outcome map from other subjects. Figure \ref{fig:proportion_simulation} shows the proportion $p(\alpha)$ (y-axis) versus the threshold $\alpha$ (x-axis) for different methods in all three scenarios. BIRD-GP dominates LR and VR in Scenarios 1 and 2. In Scenario 3 where the simulated data is in favour of VR, BIRD-GP remains robust and achieves a comparable performance to VR.

We evaluate the Monte Carlo approximate of the proposed predictor image importance measure in (\ref{eq:basis_importance_approx}) in Scenario 1, where we are able to compute the true importance measure using (\ref{eq:basis_importance}) from the data generating model. The correlation between $\widehat{\textbf{IM}}$ and \textbf{IM} is $0.953$. Supplementary S9 provides a scatter plot of $\widehat{\textbf{IM}}$ versus \textbf{IM}. This shows that the proposed Monte Carlo approximation of the proposed predictor image importance measure well reflects the actual relative importance of the predictor image importance.

\section{Analysis of fMRIs in HCP}\label{sec:hcp}
%\subsection{Overview}
In this study, we analyze fMRI data from the Human Connectome Project (HCP) 1200 release [\cite{van2012human, HCP2017}] by BIRD-GP. Previous studies have found that resting-state connectomes exhibit inter-individual differences, which can be attributed to a moderate number of connectivity components and utilized for phenotypic prediction [\cite{sripada2019}]. One type of resting-state fMRI data, fALFF, has shown to exhibit the ability to predict clinical outcomes [\cite{zhao2015bayesian, egorova2017fractional}]. On the other hand, task-evoked fMRIs demonstrate variability across individuals and serve as valuable resources for constructing predictive models of General Cognitive Ability (GCA) [\cite{spripada2020}]. An important question is to determine the extent to which the individual variability in task functional brain activity can be explained by resting-state functional brain activity alone [\cite{tavor2016task}], connectivity alone, and the combined use of both modalities. It is also of great interest to explore which one of fALFF and connectivity can provide more prediction power for task-evoked brain images. To address this inquiry, we undertake an IIR analysis, where we regress task fMRI contrast maps on fALFF images and connectivity matrices. 

We focus on two types of task fMRI contrast maps as outcome images: language task story-math contrast maps and social recognition task random-baseline contrasts maps. For each outcome type, we explore three types of predictors: fALFF alone, connectivity alone, and a combination of both modalities. Additionally, we take into account available confounders, such as age, gender, and race, in our analysis whenever possible.

\subsection{Data description and data processing}\label{sec:hcp_data}

All data analyzed in this study are from the HCP-1200 release. Data collections and analyses are performed in accordance with relevant guidelines and regulations. Figure \ref{fig:hcp_data} describes the predictors and outcomes in our study. 

The HCP language task involves two runs, with each run interleaving four blocks of story tasks and four blocks of math tasks. During the story blocks, participants are exposed to brief auditory stories adapted from Aesop’s fables (5-9 sentences). Following each story, a forced two-option question prompts the participants about the topic of the story. The math blocks entail completing addition and subtraction problems. Detailed task descriptions can be found in previous works [\cite{binder2011mapping, HCP2017}]. The HCP social recognition task consists of two runs, each comprising five video blocks (2 mental and three random in one run, three mental and two random in the other run) and five fixation blocks (15 seconds each). The videos feature 20-second clips of objects (e.g., squares, circles, triangles) either interacting in a specific manner (mental) or moving randomly (random) on the screen. After each video, participants are asked to judge whether the objects have a mental interaction, not sure, or no interaction. Detailed task information can be found in works by \citet{castelli2000movement}, \citet{wheatley2007understanding} and \citet{HCP2017}.

\begin{figure}[ht!]
    \small
    \spacingset{1}
    \centering
    \begin{subfigure}[t]{\textwidth}
        \centering
        \includegraphics[width=0.9\linewidth]{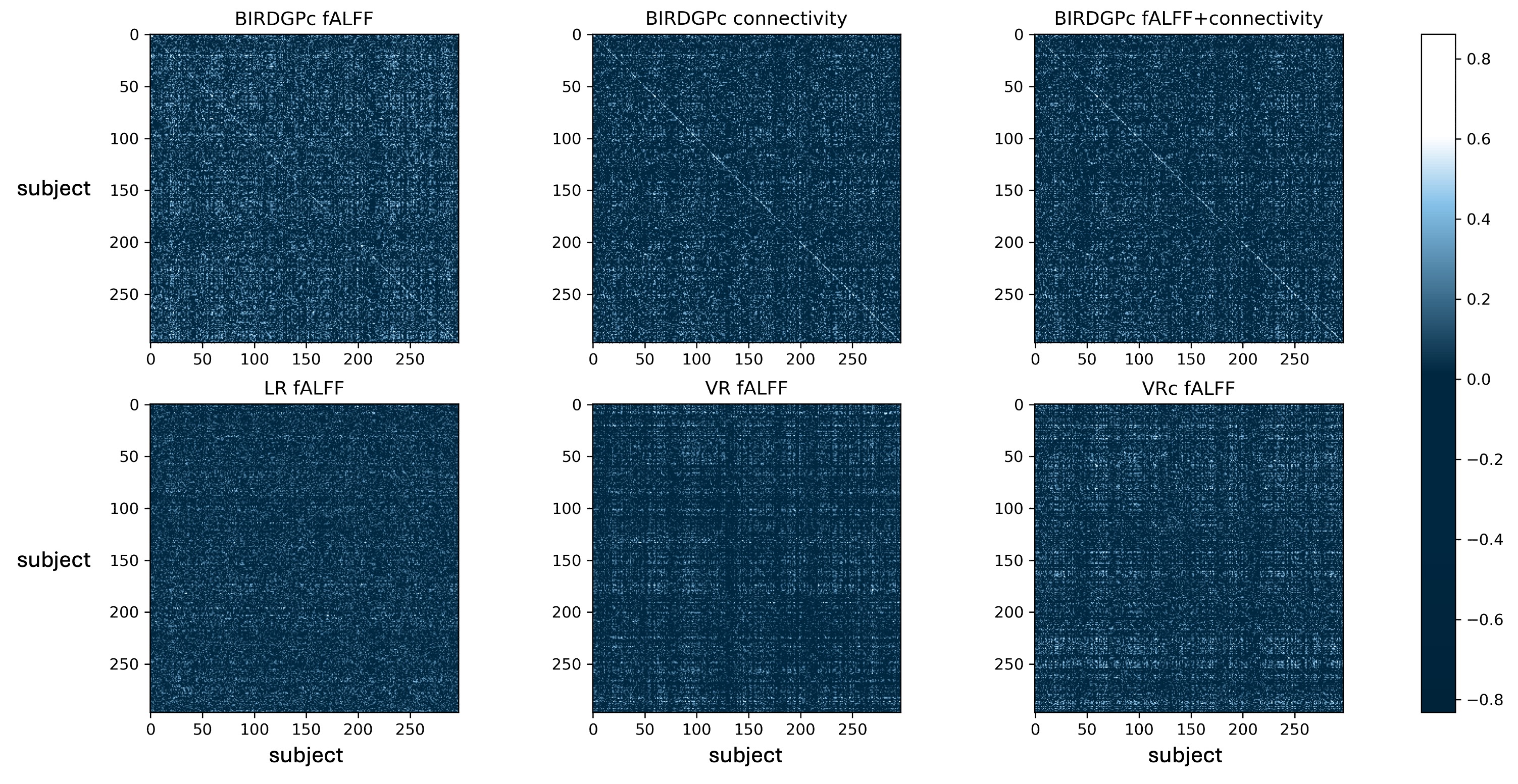}
        \caption{Language task.}
        \label{fig:correlation_lang}
    \end{subfigure}
    \vfill
    \begin{subfigure}[t]{\textwidth}
        \centering
        \includegraphics[width=0.9\linewidth]{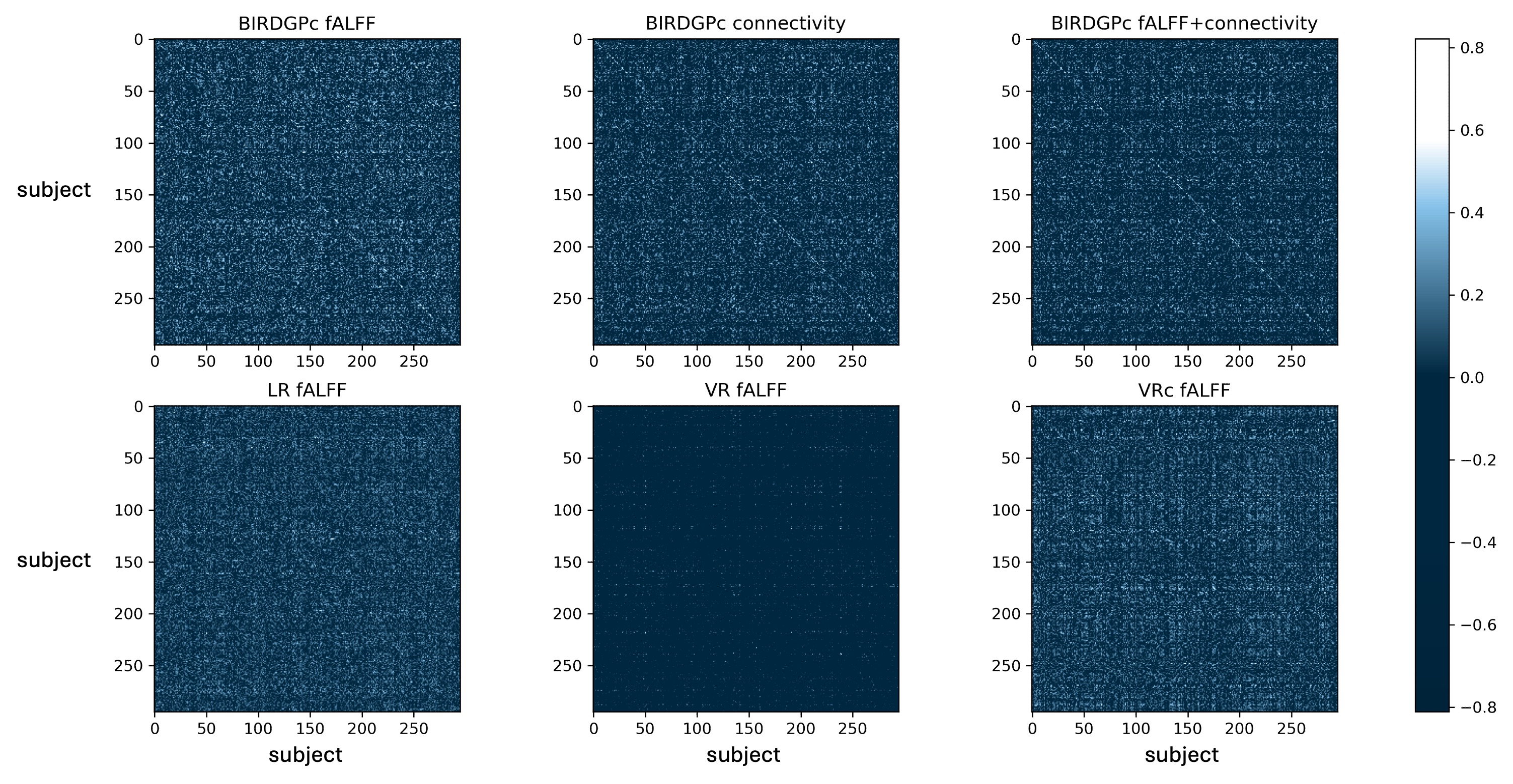}
        \caption{Social recognition task.}
        \label{fig:correlation_socirand}
    \end{subfigure}
    \caption{The prediction correlation matrix between the predicted maps (rows) and actual outcome maps (columns) for all pairs of subjects of BIRD-GP and competing methods in different task contrast maps: (a) story-math contrast from the language task and (b) random-baseline task from the social recognition task. Each entry in the matrix is the correlation between the outcome map of a subject and the same subject (diagonal), or the predicted map of another (off-diagonal). The diagonal-dominant pattern for BIRD-GP indicates BIRD-GP's prediction for any subject is more similar to the subject’s own outcome map than to other subjects’ outcome maps. In this figure, we only show subjects with more than 500 predictable activated voxels.}
    \label{fig:correlation}
\end{figure}

During the tasks, fMRI data is collected using a 32-channel head coil on a 3T Siemens Skyra scanner (TR = 720 ms, TE = 33.1 ms, 72 slices, 2 mm isotropic voxels, multiband acceleration factor = 8) with right-to-left and left-to-right phase encoding directions. Statistical parametric mapping via general linear regressions is then employed to obtain the story versus math contrast (story-math) and random versus fixation or baseline contrast (random). The data is preprocessed using the HCP minimally preprocessed pipeline [\cite{glasser2013minimal}], including gradient unwrapping, motion correction, field map distortion correction, brain-boundary based linear registration of functional to structural images, nonlinear registration to MNI152 space, and grand-mean intensity normalization. Data are then processed by a surfaced-based stream [\cite{glasser2013minimal, spripada2020}]. Both task contrast maps are obtained using the statistical parametric mapping by general linear regressions [\cite{lindquist2008statistical}]. All fALFF, story-math and random contrast map outcome images are registered into the MNI (2mm) [\cite{evans19933d}] standard brain template with a resolution of $91\times109\times91$. 

\begin{figure}[ht!]
    \small
    \spacingset{1}
    \centering
    \includegraphics[width=\linewidth]{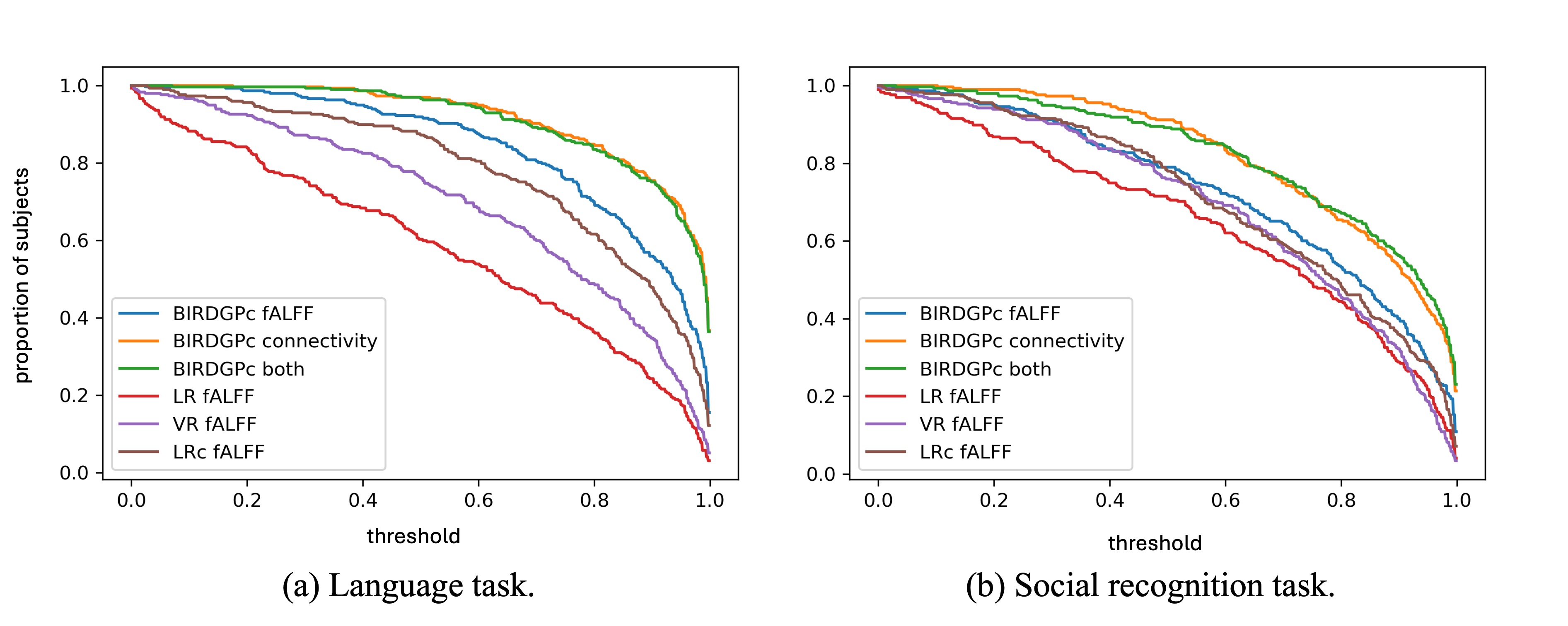}
%    \begin{subfigure}[t]{0.475\textwidth}
%        \centering
%        \includegraphics[width=\linewidth]{figures/proportion_lang_0.png}
%        \caption{Language task.}
%    \end{subfigure}
%    \hfill
%    \begin{subfigure}[t]{0.475\textwidth}
%        \centering
%        \includegraphics[width=\linewidth]{figures/proportion_socirand_0.png}
%        \caption{Social recognition task.}
%    \end{subfigure}
    \caption{The proportion of subjects $p(\alpha)$ (y-axis) that have a larger ``diagonal'' correlation than ``off-diagonal'' correlations in the same row, over different thresholds $\alpha$ (x-axis) ranging from 0-1. The ``diagonal'' correlation represents the Pearson correlation between the outcome map of a subject and the predicted map of same subject, while the ``off-diagonal'' elements are the correlation between  the predicted map of a subject and the observed map of others. The proportion measures the chance of obtaining a better prediction on the outcome map using a model prediction than using a random outcome map from other subjects. All three BIRD-GP models have a higher proportion of subjects that have a larger ``diagonal'' correlation than ``off-diagonal'' correlations at any threshold. }
    \label{fig:proportion}
\end{figure}

The resting-state connectivity correlation matrix is derived from four time-course files collected from two different fMRI sessions, each comprising 264 nodes. The 264 nodes are divided into 13 functional modules [\cite{power2011functional}], see Section S5 in the Supplementary Materials. In each session, participants complete two consecutive resting-state sessions lasting approximately 15 minutes each. The participant-specific connectivity correlation matrix is then calculated by averaging the Fisher's Z-transformed correlation matrix for the subject over the four runs.

Our analysis includes data from 714 subjects who possess all modalities (fALFF, connectivity, story-math, and random) as well as available confounders (age, gender, and race).

\subsection{Model fitting and predictive performance}\label{sec:hcp_fit}

We consider three types of predictors for our method, all adjusted for confounders: fALFF alone (BIRD-GPc fALFF), connectivity alone (BIRD-GPc connectivity), and both fALFF and connectivity (BIRD-GPc fALFF+connectivity). We compare our methods with linear regression [\cite{tavor2016task}] and voxel-wise regression [\cite{dworkin2016prevail}]. Both regression method require predictor and outcome should be from the same image space, thus can only admit fALFF as predictor. Moreover, there is no straightforward adaptation of the linear regression model to adjust for confounders. Therefore, we consider three competing models: linear regression with fALFF (LR fALFF), voxel-wise regression with fALFF (VR fALFF), and voxel-wise regression with fALFF and adjusting for confounders (VRc fALFF). %Results are summarized by brain regions annotated by the 116 automated anatomical labeling (AAL) \citep{tzourio2002automated}. 
See Section S6 in the Supplementary Materials for training details. 

We compute the prediction correlation matrix between the predicted and actual outcome maps for all pairs of subjects as in Section \ref{sec:synthetic_hcp} to assess the predictive performance of BIRD-GP and competing methods. Figure \ref{fig:correlation} shows the heatmaps of the prediction correlation matrix between the predictions and observed maps for BIRD-GP and competing methods in the two tasks. In Figure \ref{fig:correlation}, we present the heatmaps for the subjects with more than 500 voxels the predictable activated region. %where an activated voxel on a map is with the top 5\% absolute intensity values.}

In both tasks, the prediction correlation matrix $\bfC$ is noticeably diagonal-dominant for BIRD-GP, which indicates that BIRD-GP's prediction for any subject is more similar to the subject’s own outcome map than to other subjects’ outcome maps. Comparing models using only fALFF as predictors, LR fALFF and VR fALFF have barely noticeable diagonal elements, while a faint diagonal pattern stands out for VRc fALFF in both the language and social recognition tasks. The diagonal pattern becomes obvious in the correlation matrix of BIDG-GPc fALFF. This shows the advantage of BIRD-GP over competing methods. Comparing BIRD-GP with different predictors, the diagonal patterns are similar between BIRD-GPc connectivity and BIRD-GPc fALFF+connectivity, both are more visible than that of BIDG-GPc fALFF. This indicates the connectivity matrix has more prediction power than fALFF in the prediction of the task contrasts from the language and social recognition tasks, and using both fALFF and connectivity as predictors may not have a clear advantage over using connectivity alone in the the whole-brain contrast prediction of the two tasks. Comparing across the two tasks, the diagonal-dominant pattern of the correlation matrices from BIRD-GP is more noticeable in the language task than in the social recognition task. This suggests that the fALFF and connectivity are more informative in the prediction of the story-math contrast from the language task and the random-baseline contrast from the social recognition task.

To further evaluate the predictive performance, we compute the proportion of subjects $p(\alpha)$ defined in (\ref{eq:prop_subjects}) that have a larger diagonal correlation than off-diagonal correlations in the same row, over different thresholds $\alpha$ ranging from 0-1, as in Section \ref{sec:synthetic_hcp}. Figure \ref{fig:proportion} shows the proportion $p(\alpha)$ (y-axis) versus the threshold $\alpha$ (x-axis) for different methods in the two tasks. Notably, the BIRD-GP models all have a higher proportion of subjects that have a larger diagonal correlation than off-diagonal correlations at any threshold, compared to competing methods. In both tasks, BIRD-GP using connectivity alone and BIRD-GP using connectivity and fALFF have very similar performance, and is substantially better than BIRD-GP using fALFF alone. The results show the advantage of BIRD-GP over competing methods, and the advantage of including the connectivity matrix into predictors.

\begin{figure}[t!]
    \centering
    \small
    \spacingset{1}
    \includegraphics[width=0.8\linewidth]{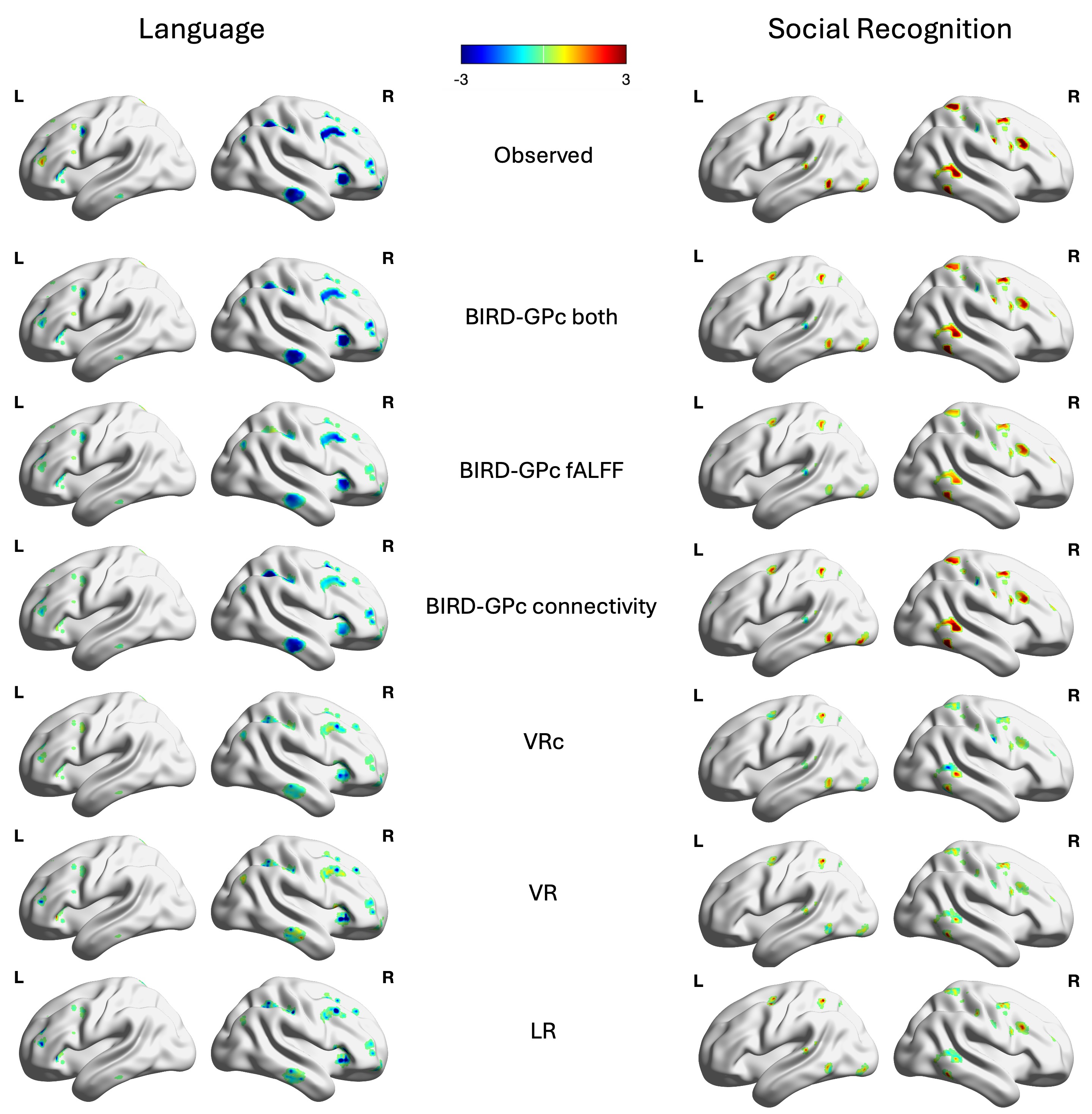}
    \caption{The observed map and the predicted map on the predictable activated region from each method for a subject with median performance in the language task (left) and the social recognition task (right).}
    \label{fig:median_subject}
\end{figure}

In Figure \ref{fig:median_subject}, we provide the observed map and the predicted map on activation voxels from each method for a subject with median performance in each task. BIRD-GP models perform better in the prediction task than other competing methods. We also examine the voxel level prediction accuracy which are reported in Section S7 in the Supplementary Material. BIRD-GP can achieve better prediction accuracy at both subject level and voxel level compared to other competing methods. We also find that the connectivity matrix is more informative than fALFF in the prediction of the story-math contrast from the language task and the random-baseline contrast from the social recognition task. The prediction power of the resting-state connectivity on the task fMRI is consistent with previous work by \citet{tavor2016task} where they found the predictor based on the resting-state functional connectivity can explain variations in task-evoked brain activity.

\begin{figure}[t]
    \centering
    \small
    \spacingset{1}
    \includegraphics[scale = 0.4]{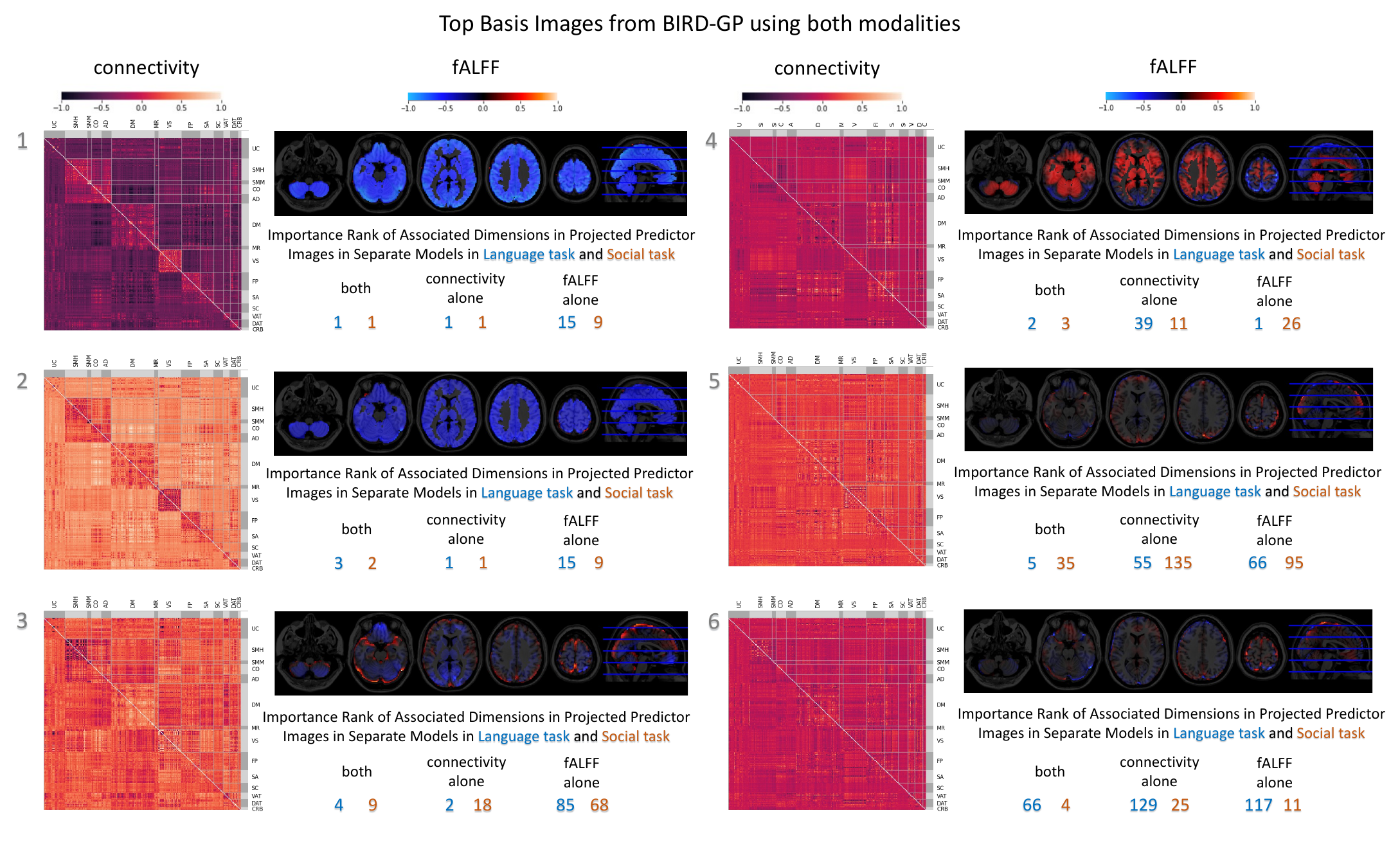}
    \caption{Six basis images associated with the most important dimensions of projected predictor images. These basis images are obtained from BIRD-GP using both fALFF and connectivity. We show the fALFF components by five slices from the axial view ($z=20, 50,80,110,140$). We also list the importance ranks of their associated projected predictor image dimensions within separate models on both the language task and social recognition task (random).}
    \label{fig:basis_importance}
\end{figure}

\subsection{Importance of the projected predictor image}

We assess the importance of all dimensions of the projected predictor image in each of the three BIRD-GP models according to (\ref{eq:basis_importance_approx}) and plot the basis images associated with most important dimensions. Figure \ref{fig:basis_importance} shows six basis images from BIRD-GPc fALFF+connectivty associated with the most important dimensions in the projected predictor image. We observe clear block patterns within functional modules in the connectivity component of basis images with high rankings. In Figure \ref{fig:basis_importance}, for each presented basis image, we also show the importance ranks of the associated dimension of the projected predictor image. As basis images produced by the three BIRD-GP models are different, we perform basis images matching across models for presentation convenience. For each basis image from BIRD-GPc fALFF+connectivity, we identify a basis image from BIRD-GPc fALFF that exhibits the highest magnitude of correlation with the fALFF component of the basis image from BIRD-GPc fALFF+connectivity. Similarly, we match a basis image from BIRD-GPc connectivity that demonstrates the largest magnitude of correlation with the connectivity component. 

In the language task, the projected predictor image dimensions associated with basis images 1, 2 and 3 are important in BIRD-GP using both modalities and BIRD-GP using connectivity alone. The projected predictor image dimension associated with basis image 4 is with high importance measure in BIRD-GP using both modalities and BIRD-GP with fALFF alone, but not important in BIRD-GP using connectivity alone. The projected predictor image dimension associated with basis image 5 is only important in BIRD-GP using both modalities, but that associated with basis image 6 is unimportant in all three models. In the social recognition tasks, the projected predictor image dimensions associated with basis images 1 and 2 are important in all three models, while those associated with basis images 3, 4 and 6 are only highly important in BIRD-GP using both modalities. The projected predictor image dimension associated with basis image 5 is not important in all three models.

To identify the functional networks most relevant to prediction, we analyzed each fALFF basis image by computing the proportion of high-intensity voxels (defined as those exceeding the 80th percentile in voxel intensity) within each network. The visual (VS) network shows the highest proportion of high-intensity voxels in basis image 1 (36\%) and basis image 3 (57\%). The dorsal attention (DAT) network is most prominent in basis image 2 (41\%) and basis image 6 (31\%). The subcortical (SC) network dominates basis image 4, accounting for 90\% of high-intensity voxels. The default mode (DM) network contributes 27\% of high-intensity voxels in basis image 5.

These results indicate that the visual, dorsal attention, and subcortical networks play key roles in predicting outcome images across both the language and social recognition tasks. This interpretation is consistent with prior literature. The visual network is actively engaged in both auditory language comprehension [\cite{ofan2007visual,seydell2023spoken,ranjan2025language}] and the perception of social interactions [\cite{pitcher2021evidence,varrier2022seeing}]. The dorsal attention network supports attentional control mechanisms during both language processing [\cite{wang2023physical}] and social cognition [\cite{callejas2014dorsal,capozzi2018attention}]. The subcortical network, including the basal ganglia and thalamus, has been shown to support language functions [\cite{ketteler2008subcortical,burgaleta2016bilingualism}] as well as social tasks [\cite{baez2013role,tanimizu2017functional}]. In contrast, the default mode network appears selectively involved in the language functions [\cite{wang2023physical}], but not in social recognition, where it is often deactivated [\cite{callejas2014dorsal}]. These findings suggest that the visual, dorsal attention, and subcortical networks provide predictive signals across both tasks, while the default mode network is specifically engaged in language tasks.

\section{Conclusion and Discussion}\label{sec:conclusion}
This paper develops BIRD-GP for IIR by flexible modeling of complex associations between the predictor and outcome images via GP-based projections and DNN. Adopting the deep kernel learning strategy, BIRD-GP can efficiently construct the basis functions that capture the detailed characteristics of both the predictor and outcome images very well. Compared with other state-of-the-art methods, BIRD-GP can greatly reduce the number of model parameters, improve the prediction accuracy for IIR tasks and provide a set of basis images, leading to better interpretations of the model fitting. Our analysis of the HCP fMRI data by BIRD-GP reveals that the connectivity matrix has much more predictive power than fALFF on contrast maps from two HCP tasks, and combining both modalities achieve the best performance. We also identify the important functional networks in the prediction of both the language task story-math contrast and the social recognition random-baseline contrast.

Several strategies can be adopted for potential improvement on BIRD-GP's training time. First, refitting basis coefficients after orthogonalization could be parallelized across subjects, while we currently use serial computation. Additionally, we use Bayesian linear regression for refitting coefficients after orthogonalization, though ordinary linear regression (or elastic net regression) could be employed to significantly reduce computation time. For instance, in the Fashion MNIST experiment with 1000 training samples, using linear regression could reduce the refitting time to under 10 seconds or less than one second with full parallelization, bringing down the total training time to approximately 250 seconds. Lastly, kernel learning could be accelerated by leveraging transfer learning on datasets with similar characteristics, enabling the reuse and fine-tuning of pre-learned basis functions. 

\section*{Funding}
This work was partially supported by NIH grants (R01DA048993, R01MH10 5561) and NSF grant IIS2123777.

\bibliographystyle{apalike}
\bibliography{ref}

\end{document}